\documentclass[aps,prd]{revtex4}

\usepackage{amsbsy}
\usepackage{amssymb}
\usepackage{amsmath}
\usepackage{graphicx}

\newcommand{\unit}[1]{%
    \ensuremath{\, \mathrm{#1}}}

\def\jnl@style{\it}

\def\aaref@jnl#1{{\jnl@style#1}}

\def\aj{\aaref@jnl{AJ}}                   
\def\apj{\aaref@jnl{ApJ}}                 
\def\apjl{\aaref@jnl{ApJ}}                
\def\apjs{\aaref@jnl{ApJS}}               
\def\apss{\aaref@jnl{Ap\&SS}}             
\def\aap{\aaref@jnl{A\&A}}                
\def\aapr{\aaref@jnl{A\&A~Rev.}}          
\def\aaps{\aaref@jnl{A\&AS}}              
\def\mnras{\aaref@jnl{MNRAS}}             
\def\prc{\aaref@jnl{Phys.~Rev.~C}}        
\def\prd{\aaref@jnl{Phys.~Rev.~D}}        
\def\pre{\aaref@jnl{Phys.~Rev.~E}}        
\def\prl{\aaref@jnl{Phys.~Rev.~Lett.}}    
\def\qjras{\aaref@jnl{QJRAS}}             
\def\skytel{\aaref@jnl{S\&T}}             
\def\ssr{\aaref@jnl{Space~Sci.~Rev.}}     
\def\zap{\aaref@jnl{ZAp}}                 
\def\nat{\aaref@jnl{Nature}}              
\def\aplett{\aaref@jnl{Astrophys.~Lett.}} 
\def\apspr{\aaref@jnl{Astrophys.~Space~Phys.~Res.}} 
\def\physrep{\aaref@jnl{Phys.~Rep.}}      
\def\physscr{\aaref@jnl{Phys.~Scr}}       
\def\commat{\aaref@jnl{Comm.~Math.~Phys.}}      
\def\science{\aaref@jnl{Science}}       
\def\cqg{\aaref@jnl{Class.~Quantum Gravity}}        
\def\jpcs{\aaref@jnl{JPCS}}                 
\def\ijmp{\aaref@jnl{Int.~J.~Mod.~Phys.}}           
\def\npa{\aaref@jnl{Nucl.~Phys.~A}}        

\newcommand{\diff}{\mathop{}\!\mathrm{d}}

\begin{document}

\title{Seismology of adolescent neutron stars: Accounting for thermal effects
and crust elasticity}

\author{C.\,J.\,Kr\"uger, W.\,C.\,G.\,Ho, N.\,Andersson}

\affiliation{
Mathematical Sciences and STAG Research Centre, University of Southampton,
Southampton SO17 1BJ, United Kingdom
}

\begin{abstract}
We study  the oscillations of relativistic stars, incorporating key physics
associated with internal composition, thermal gradients and crust
elasticity.  Our aim is to develop a formalism  which is able to account for
the state-of-the-art understanding of the complex physics associated with
these systems.  As a first step, we build models using a modern equation of
state including composition gradients and density discontinuities associated
with internal phase-transitions (like the crust-core transition and the  point
where muons first appear in the core).  In order to understand the nature of
the oscillation spectrum, we carry out cooling simulations to provide
realistic snapshots of the temperature distribution in the interior as the
star evolves through adolescence. The associated thermal pressure is
incorporated in the perturbation analysis, and we discuss the presence of
$g$-modes arising as a result of thermal effects. We also consider interface
modes due to phase-transitions and the  gradual formation of the star's crust
and the emergence of a set of shear modes.
\end{abstract}

\maketitle

\section{Introduction}

Neutron stars are seismically complex, with particular classes of oscillation
modes associated with specific parts of the involved physics. This makes the
construction of a truly realistic model of the star's oscillation spectrum a
complicated task. Nevertheless, we are reaching the point where much of the
physics involved is understood at the level required to build moderately
realistic models.  The aim of this work is thus quite natural; we want to take
important steps towards realism by accounting for processes that are relevant
as a neutron star matures. We achieve this by tracking the cooling of a given
star from a minute or so after birth through the first several hundred years,
paying particular attention to the changes in the thermal pressure and the
formation of that star's elastic crust.  The output from our state-of-the-art
cooling code provides input for the seismology analysis. The results provide a
sequence of snapshots of how the star's oscillation spectrum evolves as the
star ages. This is an important advance on previous work in this area.

The problem of relativistic star seismology has been considered since
pioneering work by  Thorne \& Campolattaro in the late 1960s
\cite{thorne_campolattaro_1967}. Like much of the subsequent work, the initial
focus was on the mathematical formulation of the problem with the complex
supranuclear physics playing a secondary role. Thus, the bulk of the
literature is focussed on perfect fluid stars, often without any consideration
of the interior composition and state of matter.  In fact, many studies have
made use of rather ad hoc polytropic models which only capture the rough
properties of a realistic equation of state. Nevertheless, this has led to an
understanding of the basic nature of the stellar spectrum, like the
fundamental $f$-mode and the pressure restored $p$-modes
\cite{lindblom_detweiler_1983, detweiler_lindblom_1985}, and some discoveries,
like the existence of the gravitational-wave $w$-modes,
\cite{kokkotas_schutz_1992}. However, the understanding of the fine details of
the problem has so far been developed one piece at a time. Gravity $g$-modes
arising because of composition gradients have been considered \cite{finn_1986,
finn_1987, strohmayer_1993, miniutti_etal_2003}, the crust elasticity has been
accounted for (especially for shear modes) \cite{schumaker_thorne_1983}, the
roles of core superfluidity \cite{comer_etal_1999, andersson_etal_2002,
lin_etal_2008} and the star's magnetic field \cite{sotani_etal_2008,
colaiuda_etal_2009, gabler_etal_2012, gabler_etal_2013} have been investigated
and oscillations of proto-neutron stars and their
detectability~\cite{ferrari_etal_2003} have been studied, as well. In the last
decade, much
attention has been focussed on the Coriolis restored
$r$-modes~\cite{andersson_1998, andersson_kokkotas_2001, lockitch_etal_2001,
haskell_etal_2009}, which are relevant because they may be driven unstable
due to the gravitational radiation they generate. In this context, much of the
complex microphysics related to dissipative processes has been discussed,
progressing our understanding of the involved processes considerably.

Neutron star oscillations may impact on a range of observations, involving in
particular radio and X-ray timing and gravitational waves. At the present
time, the most promising connection between observations and theory is
provided by the observed quasiperiodic oscillations seen in the X-ray tail of
magnetar giant flares \cite{watts_strohmayer_2006, samuelsson_andersson_2007}.
The inferred oscillation frequencies match those expected for the elastic
crust reasonably well, and provide the first credible example of actual
neutron star asteroseismology.  There are, of course, issues to be resolved,
especially concerning the dynamical role of the interior magnetic field in
these systems \cite{colaiuda_kokkotas_2011}. A more indirect example comes
from X-ray timing of fast spinning accreting neutron stars in Low-Mass X-ray
Binaries \cite{strohmayer_mahmoodifar_2014, andersson_etal_2014}.  The spin of
these stars appears to be limited by some process. One of the leading
contenders for the underlying mechanism is the $r$-mode instability which leads
to the star losing angular momentum at a rate that could balance the spin-up
due to accretion \cite{andersson_etal_1999, bondarescu_etal_2007}. Finally,
as we are getting closer to the advent of gravitational-wave astronomy, there
has been searches for neutron star oscillations from a range of systems
\cite{ligo_search1, ligo_search2, ligo_search3, ligo_search4}. At present,
these studies may only be providing relatively uninteresting
upper limits on the possible signals. With an advanced generation of detectors
coming online in the next few years, this could turn to actual detections. 

The existing body of work allows us to piece together a picture of neutron star dynamics. This understanding should be valid provided the star is in a regime where the different ingredients remain distinct. Unfortunately, this is unlikely to be the case. Hence, there is a pressing need to develop a new generation of models that account for as much of the relevant physics as possible. The present work should be seen in that context. 

The physics associated with realistic neutron star dynamics is daunting. Many
aspects are  rather poorly understood, in particular concerning the deep core
(at several times the nuclear saturation density). Nevertheless, a focussed
effort on this problem is timely. First of all, gravitational-wave astronomy
should become reality in the next few years. This promises to provide us with
observational data (perhaps most likely from the merger of compact binaries)
that need to be matched against the best possible theoretical models.
Secondly, our understanding of the principles associated with neutron star
superfluidity has improved considerably in the last decade. In particular, we
have constraints on the superfluid parameters (the superfluid pairing gaps or,
equivalently, the critical temperatures) from the observed real-time cooling
of the remnant in Cassiopeia A \cite{shternin_etal_2011}. In order to be able
to make productive use of future observational results, we need to improve our
level of modelling. 

The basic requirement for progress on this problem is a realistic equation of
state which accounts for the two-fluid nature of a neutron star's core,
including information about the superfluid gap energies and the entrainment
parameters. We also need a comprehensive formalism for modelling dynamics
which accounts for all the desired parameters. Progress on the first of these
issues was recently made by Chamel \cite{chamel_2008}, who provided the first
ever consistent equation of state including entrainment. The second problem
has been explored in basic models \cite{andersson_etal_2013}, to the
point where there are no technical stumbling blocks preventing more realistic
studies. 

In this work we  take the first steps towards true realism, starting from the
classic fluid formalism of Detweiler \& Lindblom
\cite{detweiler_lindblom_1985} and extending it to account for density
discontinuities associated with distinct phase-transitions, interior
composition gradients, thermal pressure, and the elastic crust which will form
and grow in thickness as the neutron star cools. We use this model  to study
the evolution of the star's oscillation modes as it matures. To do this we
couple the oscillation mode calculation to the long-term cooling of the star.
At different points in the thermal evolution we output the temperature profile
and feed it into the mode calculation.  The results of this exercise shed
light on the influence of thermal effects on the various oscillation modes of
the star. As expected, we find that the fundamental $f$-mode and the various
$p$-modes are only weakly affected by the changes in temperature. Meanwhile, the
gravity $g$-mode spectrum changes completely as the thermal pressure weakens and
we demonstrate how only a few interface modes (arising from density
``discontinuities") remain (in the considered frequency range above 10~Hz or
so)  when the star reaches maturity. Finally, the presence of the elastic
crust enriches the spectrum by shear modes, which also evolve as the star
cools and the crust region grows. 

This paper is the first in a series where we aim to build a truly realistic
model of a dynamical neutron star. Our initial focus is on issues relating to
composition and thermal effects. This will lead naturally on to the role of
superfluidity, which will be considered in a follow-up paper. The basic reason
for the division is that the problems we consider here can be modelled within
    the standard single-fluid framework, while superfluidity adds  degrees of
    freedom that make the problem richer (and obviously more complicated).
    The same is true for the star's magnetic field, which would eventually
    have to be included in the model (involving issues associated with the
    presence of a superconducting core \cite{glampedakis_etal_2011}).

The layout of the paper is as follows: Section \ref{sec:ns_model} is devoted
to the physics which we take into account for the background model; in Section
\ref{sec:perturbations}, we discuss the perturbation equations and explain how
these are modified to account for additional physics (the perturbation
equations for the elastic crust are given in
Appendix \ref{sec:app_eq_crust_polar}). A general numerical strategy for
solving the perturbation equations in the interior of a multi-layered star is
presented in Section \ref{sec:num_strategy} and
Appendix \ref{sec:app_numerics}; Section \ref{sec:num_strategy} also covers
the reasoning for a new set of equations for perturbations of the perfect
fluid, which aids in solving the perturbation problem at low frequencies; the
actual equations can be found in Appendix~\ref{sec:app_eq_low_freq}. Finally,
Section \ref{sec:results} contains the results and Section \ref{sec:summary}
summarises our work.  Unless stated otherwise, we use units in which $G=c=1$
and Misner, Thorne and Wheeler (MTW) \cite{MTW1973} conventions throughout the
paper.

\section{Key aspects of the Model}
\label{sec:ns_model}

We aim to build a state-of-the-art neutron star model
and track changes in its dynamics, represented by non-radial modes of
oscillation, as the star evolves from just after birth into adolescence. The
problem involves a number of unknowns, ranging from the bulk equation of state
(e.g. pressure versus density in the deep core) to details of the microphysics
(like the reactions that dictate the thermal evolution). In order to make
progress, and avoid unnecessary confusion, it is necessary to make choices 
from the very beginning. 

Given the uncertainties involved, much work on neutron-star seismology has
considered a range of equations of state, aiming to establish to what extent
observations can be used to distinguish between different models
\cite{andersson_kokkotas_1998}. This is a useful strategy as it provides a
foundation for future efforts to carry out the asteroseismology programme.
However, it is not a practical approach for our present purposes. The many
different options involved would cause undue confusion. Instead, we will
assume that the physics input is known (even though it clearly is not!) at the
required level of precision. This allows us to construct a unique sequence of
stellar models. In order to be considered ``realistic'',  this sequence must
satisfy all current constraints from observations. In particular, the model
must allow the maximum mass to be above 2$M_\odot$
\cite{demorest_etal_2010,antoniadis_etal_2013} while the radius of a typical
stellar model should lie in the range inferred from X-ray burst sources. For
the latter, we use the result from \cite{steiner_etal_2010}, i.e. a radius in
the range of $11.5 \pm 1.2\unit{km}$. Fortunately, the SLy4 equation of state
\cite{chamel_2008,haensel_potekhin_2004} satisfies these criteria. This model
also suits our purposes by involving the simplest reasonable composition. The
star's core contains only neutrons, protons, electrons and muons; there are no
``exotic'' components like hyperons or deconfined quarks.

Having fixed the equation of state, we still need to narrow the focus.  We
will consider a sample neutron star. The aim is to track the changes in the
oscillation spectrum as this star ages.  The particular star we consider has
central density $\rho_c = 1 \times 10^{15} \unit{g} \unit{cm}^{-3}$, which
leads to a radius of $R = 11.77\unit{km}$ and a mass of $M =
1.45\,M_{\odot}$. This model star cools mainly due to
modified Urca processes \cite{ho_etal_2012}.

\subsection{The Background Configuration}
\label{sec:background}

In order to solve the seismology problem, we first of all need a background
solution. The construction of such a configuration is standard. The
equilibrium of a non-rotating star with an unstrained crust is described by a
static, spherically symmetric spacetime with metric $g_{ab}$ that leads to the
line element
\begin{equation}
    ds^2 = - e^{\nu} dt^2 + e^{\lambda} dr^2 + r^2 d\theta^2
            + r^2 \sin^2\theta d\phi^2
\end{equation}
and the standard perfect fluid energy-momentum tensor
\begin{equation}
    T_{ab} = (\rho + p) u_{a} u_{b} + p g_{ab},
\end{equation}
where $\rho$ is the energy density and $p$ the pressure. The fluid
four-velocity is given by
\begin{equation}
    u^a = e^{-\nu/2} t^a,
\end{equation}
with $t^a = (\partial_t)^a$ the timelike Killing vector of the
spacetime.

At finite temperatures, the pressure $p$ is given by a two-parameter equation
of state~\footnote{Note that the problem changes if we want to account for
superfluid components.} (EoS henceforth) which depends on the energy density
$\rho$ and the temperature $T$ (or the entropy density  $s$). In the first few
minutes after the star's birth, the thermal pressure is considerable and the
star is puffed up to roughly twice its final radius. In order to simplify the
analysis, we will not consider this stage. A study that complements ours in
this respect has already been carried out
\cite{burgio_etal_2011}.  In the
following we consider models that are sufficiently cold that we can neglect
the  thermal pressure for the background configuration (but not for the
perturbations!). In effect, this means that the star has to be colder than
$10^{10}$~K or so. The background EoS can then be given in one-parameter form
\begin{equation}
    p = p(\rho).
\end{equation}

The Einstein equations lead to the well-known
Tolman-Oppenheimer-Volkoff (TOV) equations
\begin{align}
    \lambda'
        & = \frac{1-e^{\lambda}}{r} + 8\pi r e^{\lambda} \rho,
                                                    \label{eq:tov_lambda} \\
    \nu'
        & = \frac{e^{\lambda}-1}{r} + 8\pi r e^{\lambda} p,
                                                    \label{eq:tov_nu} \\
    p'
        & = - \frac{1}{2}(\rho + p) \nu',
                                                    \label{eq:tov_p}
\end{align}
where a prime denotes a derivative with respect to $r$.
We also define the mass inside radius $r$ to be 
\begin{equation}
    M(r) = \frac{1}{2} r \left( 1 - e^{-\lambda} \right).
                                                    \label{eq:def_M}
\end{equation}
Finally, the adiabatic index of the background configuration, which is
assumed to be in $\beta$-equilibrium, is
\begin{equation}
    \label{eq:def_gamma_0}
    \gamma_0 = \frac{\rho + p}{p} \frac{dp}{d\rho}.
\end{equation}

\subsection{The Equation of State}
\label{ssec:eos}

To model the star's core, we use the numerical fit to the SLy4 EoS
proposed by Chamel \cite{chamel_2008,haensel_potekhin_2004}. This is
a zero temperature EoS which accounts for the presence of a mixture of
superfluid neutrons and superconducting protons in the core, and includes
the entrainment effect. Estimates of the associated parameters are key to
future developments of our model so it is natural to focus on this particular
EoS already at this stage.

As a neutron star matures, the state of matter changes. The outer layers
freeze to form the elastic crust and the core becomes a
superfluid/superconducting mixture.  We will account for the former, but
postpone consideration of the latter. However, provided that the elastic crust
has time to relax we do not have to account for the presence of strains in the
background model.  This effectively means that we are still dealing with
pressure only as a function of the density. In the inner crust (up to the
pressure $p_m = 3.755 \times 10^{32} \unit{dyn} \unit{cm}^{-2}$), our model
makes use of the (simplified) EoS proposed by Douchin \& Haensel (DH)
\cite{douchin_haensel_2001}; the outer crust of our star is modelled by the
EoS proposed by Haensel \& Pichon \cite{haensel_pichon_1994}, however, we use
the refined values provided by Samuelsson \cite{samuelsson_thesis}.

The matching of the two chosen EoS at the pressure $p_m$  results in a
small, artificial density discontinuity of $\Delta \rho / \rho_+ \approx
3.0\,\%$ at the crust-core transition (where $\rho_+$ is the density of the
crustal matter at the transition). This piecing together of different EoSs
might seem a bit ad-hoc; however, our main interest is in the development of
the technology required to solve the problem.  We aim to provide a proof of
principle, not a truly accurate spectrum of a real neutron star. Of course,
the computational technology is such that, if we are provided with the true
EoS the analysis could proceed more or less in a plug-and-play fashion.
Density discontinuities, such as the artificial one mentioned here are not a
numerical problem but an expected feature. Within the crust of a neutron star where
matter is formed from ions, there are several first order phase-transitions
which come with a sudden change in density. The DH EoS accounts for 12 such
phase-transitions and the associated density discontinuities are of similar
size as our artificial one: $\Delta \rho / \rho_+$ varies between
$2.0-4.3\,\%$ (where $\rho_+$ denotes the lower density of a discontinuity).

\subsection{Thermal Evolution}
\label{ssec:cooling}

\begin{figure}[ht]
    \centering
    \includegraphics[clip=true,width=0.6\textwidth]{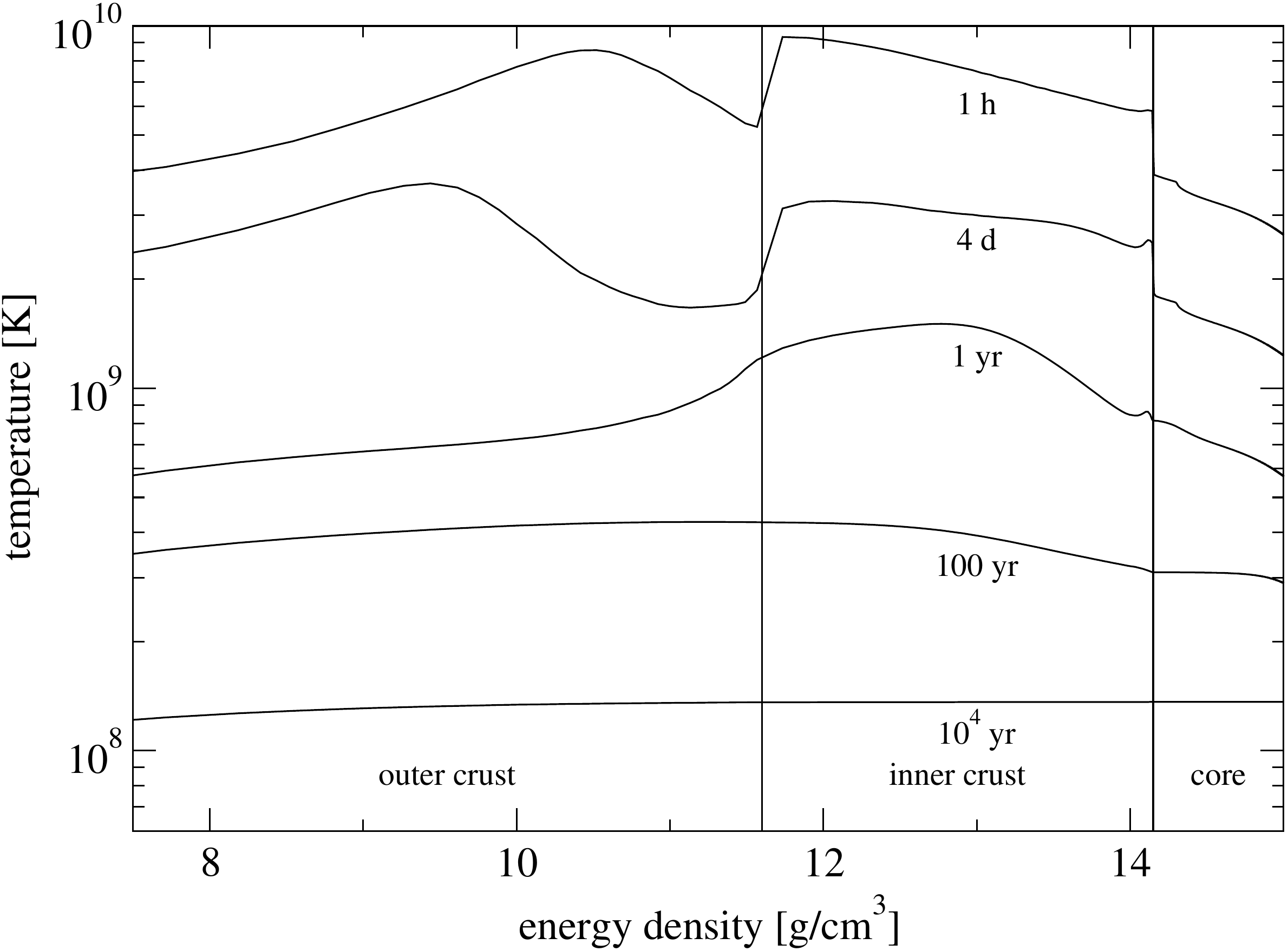}
    \caption{The thermal evolution of our neutron star model with $M =
    1.45\,M_{\odot}$. As is apparent from this graph, the neutron star is
    nearly isothermal after 100 years (without heat sources).}
    \label{fig:cooling}
\end{figure}

In order to quantify the thermal effects on the star's oscillation spectrum, we carry out cooling simulations from $T \approx
10^{10}\unit{K}$, at which point the thermal pressure can be neglected compared to the static pressure.
Starting from a uniform initial temperature profile (this is artificial but filters out of the system very rapidly), we evolve the interior
temperature $T(r,t)$ of the neutron star using the relativistic equations of
energy balance and heat flux \cite{ho_etal_2012}
\begin{align}
    \frac{ e^{-\lambda/2-\nu} }{4 \pi r^2}
        \frac{\diff}{\diff r} \left( e^{\nu} L_r \right)
        & = - e^{-\nu/2} C \frac{\partial T}{\partial t}
          - \varepsilon_{\nu},
                                                        \label{eq:temp1} \\
    \frac{ L_r }{ 4 \pi r^2 }
        & = - e^{ -(\lambda+\nu)/2 } K
          \frac{\partial}{\partial r} \left( e^{\nu/2}  T \right),
\end{align}
where $L_r$ is the luminosity at radius $r$, $C$ is the heat capacity,
$\epsilon_{\nu}$ is the neutrino emissivity and $K$ is the thermal
conductivity. Note that we do not include internal heat sources,
which would appear as a further source term in Equation \eqref{eq:temp1};
without these, neutron stars become essentially isothermal after about 100
years, as is apparent in Figure~\ref{fig:cooling} where we show the thermal
evolution of our chosen neutron star. 

In order to provide a picture of what is involved, let us
list the ingredients going into this calculation (for a detailed discussion we
refer the reader to Ho et~al.~\cite{ho_etal_2012}). The total heat capacity $C$ is the sum of the partial heat capacities. In the
core, these are due to neutrons, protons, electrons and muons whereas in the
crust there are contributions from  free neutrons, ions and electrons
\cite{vanriper_1991,ho_etal_2012}.

The neutrino emissivity $\epsilon_{\nu}$ has several contributors as well and
must be treated differently in the core and in the crust.  In the core, we
account for two classes of neutrino emission processes.  First, we have the
so-called modified Urca processes; we consider both the neutron and the proton
branch. The direct Urca process, however, does not contribute since it occurs
only at densities higher than those reached by the stellar model we consider
here, for which the central density is $\rho_c = 1 \times
10^{15}\unit{g}\unit{cm^{-3}}$.  Second, we consider bremsstrahlung in which
neutrino-antineutrino pairs are produced. In particular, we use emissivities
due to neutron-neutron, neutron-proton and proton-proton scattering. The
emissivities we  use were calculated by Yakovlev
et~al.~\cite{yakovlev_etal_1999} and Page et~al.~\cite{page_etal_2004}. In the
crust, bremsstrahlung is still an important process and we consider
electron-nucleon, neutron-neutron and neutron-nucleon scattering.
Additionally, we consider the two processes of plasmon decay and
electron-positron pair annihilation which could both in principle occur in the
core but are much more efficient in low density regions like the crust. We
take the emissivities for these processes from Yakovlev
et~al.~\cite{yakovlev_AK_KL_1999,yakovlev_etal_2001}. For the thermal
conductivity of the core, we sum the contributions due to neutrons, electrons
and muons \cite{flowers_itoh_1979,flowers_itoh_1981}. We take the results of
Baiko et~al.~\cite{baiko_etal_2001} for the neutron thermal conductivity and
the results of Shternin \& Yakovlev \cite{shternin_yakovlev_2007} for the
electron and muon thermal conductivities. For the crust, we use {\sc
conduct08} \cite{conduct08}, which implements the latest advances in
calculating thermal conductivities
\cite{potekhin_etal_1999,cassisi_etal_2007,chugunov_haensel_2007}.

The outer layers --- the envelope --- of the crust serve as a heat blanket and
can support large temperature gradients near the surface of the star
\cite{gudmundsson_etal_1982}. The relationship between the temperature at the
bottom of the envelope $T_{\text{env}}$ and the effective temperature of the
photosphere $T_s$ depends on the composition of the envelope. We consider
envelopes consisting of either iron or light elements and use relations
between $T_{\text{env}}$ and $T_s$ given by Potekhin
et~al.~\cite{potekhin_etal_2003}.

As is apparent from Figure~\ref{fig:cooling}, the core of the neutron star
cools more quickly than the crust in the very early stages of its life. This
is due to stronger neutrino emission in the core and hence the crust is
generally hotter than the core. In these very early stages, thermal
conductivity does not play a major role in the cooling evolution and the
cooling of the core and the crust can be considered more or less decoupled;
this leads to the rather big jumps in temperature at the crust-core
transition, see Figure~\ref{fig:cooling}. The jump between the outer and inner crust is due to the neutron
drip. After about one year or so, the star has cooled considerably,
so that the cooling mechanisms become less efficient and conductivity
will smooth the temperature profile throughout the star, finally bringing
it into an isothermal state after about 100 years.

\subsection{Crust Formation}
\label{ssec:crust_formation}

As one of our aims is to study shear modes associated with the crust
elasticity, we have to specify the region in which the crust may sustain shear
stresses. That is, we need to quantify the solid region.  The standard
idealised model for the crust of a neutron star is a one-component plasma, in
which ions of charge $Ze$, number density $n$ and temperature $T$ are free to
move. The thermodynamics of such a plasma can be described by the
dimensionless Coulomb coupling parameter
\begin{equation}
    \Gamma = \frac{ (Ze)^2 }{ akT },
\end{equation}
where $a=(4 \pi n/3)^{-1/3}$ denotes the inter-ion spacing and
$k$ is the Boltzmann constant. Calculations by Farouki \& Hamaguchi
\cite{farouki_hamaguchi_1993} show that the crust crystallises as $\Gamma$
increases (due to cooling) above $\Gamma_m \approx 173$. The DH crust EoS (see
Section \ref{ssec:eos}) we are using provides us with all parameters required
to quantify the crust freezing.  Given  the temperature profiles from our
thermal evolution we calculate $\Gamma$ and assume the crust to be elastic in
the region in which $\Gamma \ge \Gamma_m$. The quantity $\Gamma_m$ is
associated 
with uncertainties of a few percent; a slightly different value will affect
the thickness of the crust and thereby the frequencies of the shear modes.
Similar to the choice of the EoS, we do not further investigate these effects
in this study and take the value of $\Gamma_m = 173$ as given. It can easily
be adjusted once more accurate data are available.

\begin{figure}[ht]
    \centering
    \includegraphics[clip=true,width=0.6\textwidth]{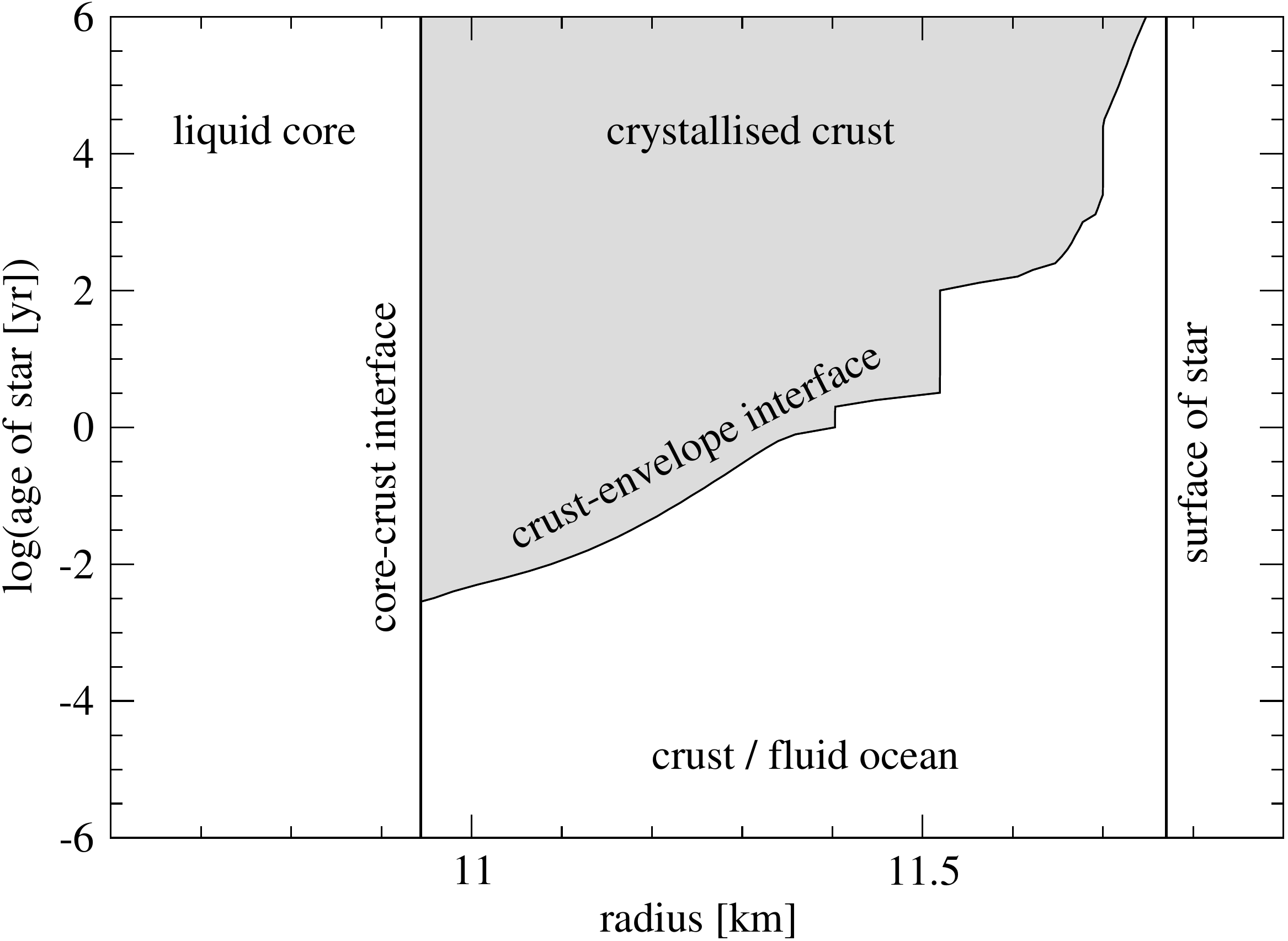}
    \caption{The formation of the solid crust over time. We show only the
    outer layer of our model star. The area where
    the crust is crystallised is shaded in grey; this region is calculated
    using a sharp threshold of $\Gamma > 173$ (see text).}
    \label{fig:crust_formation}
\end{figure}

In Figure~\ref{fig:crust_formation}, we show how the crust region evolves as
the star cools. The crust of the neutron star stays liquid for a bit more than
a day; then the crust starts to crystallize at the core-crust interface.
Within the first three years the crust gains quickly in width up to about
580m. After about hundred years without any significant further
crystallisation, the temperature has dropped below the melting temperature
also in the outer layers of the crust and the elastic crust eventually
extends nearly to the surface of the star.

\subsection{Perturbations}
\label{sec:perturbations}

Our model accounts for several ``restoring forces'' expected to be present in
a real neutron star, each of which results in a certain, more or less distinct
class of modes in the stellar spectrum. The model considered here obviously
has the standard $f$-mode (due to the presence of the star's surface) and
acoustic $p$-modes, restored by the pressure. In addition we have  composition
gradients (leading to composition $g$-modes), a finite temperature (thermal
$g$-modes), an elastic crust ($s$-modes) and density discontinuities
($i$-modes). In this section we  discuss how the relevant physics enters the
perturbation equations.

Before we describe the additional physics, we briefly mention basic
definitions. We define the Lagragian displacement vector $\xi^a$ to be
\begin{align}
    \label{eq:def_xi}
    \xi^t
        & = 0, \\
    \xi^r
        & = r^l e^{-\lambda/2} \frac{W}{r} P_l e^{i \omega t}, \\
    \xi^{\theta}
        & = - r^l \frac{V}{r^2} \frac{\diff P_l}{\diff \theta} e^{i \omega t}, \\
    \xi^{\phi}
        & = 0,
\end{align}
where $W$ and $V$ are functions of $r$ and $P_l$ are the Legendre polynomials
which account for the angular dependence. The velocity perturbation is then
given by
\begin{equation}
    \delta u^a = q^a_b \left( u^c \nabla_c \xi^b - \xi^c \nabla_c u^b \right)
        + \frac{1}{2} u^a u^b u^c h_{bc},
\end{equation}
with the projection tensor $q^{ab} = g^{ab} + u^a u^b$ and $h_{ab}$ the
perturbed metric as given later (see Equation \eqref{eq:def_h}).

\subsubsection{Stratification}

Let us first consider the effects of internal stratification on a given
perturbed fluid element. As the fluid moves, weak interaction processes try to
adjust the composition to the surrounding matter and restore
$\beta$-equilibrium.  However, as Reisenegger \& Goldreich
\cite{reisenegger_goldreich_1992} have argued, the timescale of the relevant
weak interaction processes is much longer than the typical oscillation period
and thus perturbed fluid elements are not able to equilibrate during one
oscillation. In fact, it is usually appropriate to assume that  the
composition of a perturbed fluid element is frozen. This means that  the
adiabatic index $\gamma$ of the perturbed fluid is
\begin{equation}
    \label{eq:def_gamma}
    \gamma
        = \frac{\rho + p}{p}
          \left( \frac{\partial p}{\partial \rho} \right)_{n,S},
\end{equation}
where $S$ is the entropy per baryon. We will refer to this as the \emph{slow
reaction} limit.

In order to understand the different classes of $g$-modes, it is useful to introduce the Schwarzschild discriminant
\begin{equation}
    \label{eq:def_schw_discr}
    A = \frac{dp}{dr} - \frac{\gamma p}{\rho + p} \frac{d\rho}{dr}
        = \left( 1 - \frac{\gamma}{\gamma_0} \right) \frac{dp}{dr},
\end{equation}
which determines the stability of a pulsating star against convection
\cite{detweiler_ipser_1973}. Stability requires that  $A \ge 0$
throughout the star. For our particular EoS this condition holds
as we have $\gamma \ge \gamma_0$ everywhere. Note that we have used
Equation~\eqref{eq:def_gamma_0} in order to achieve the last equality in
Equation~\eqref{eq:def_schw_discr}. Composition $g$-modes arise as the
buoyancy associated with $A$ provides a restoring force for small
fluid-element deviations from equilibrium.

It may, of course, be the case that the dynamics is much slower than the weak
interactions.  In the  limit of \emph{fast reactions} the  interactions are
sufficiently fast to fully adjust the perturbed fluid element's composition
such that it stays in $\beta$-equilibrium at all times. Such a star is called
barotropic and the adiabatic indices, $\gamma$ and $\gamma_0$, then coincide.
As a consequence, the Schwarzschild discriminant vanishes, $A = 0$. In this
case, perturbed fluid elements are not affected by buoyancy and the $g$-modes
due to stratification are absent from the stellar spectrum.

\subsubsection{Thermal Pressure}

Acting in a fashion similar to that of composition gradients, the presence of
a finite temperature leads to a thermal pressure that influences the fluid
dynamics. This effect can be important in young neutron stars. Since our
EoS models zero temperature physics, we will account for thermal
effects by  adding the thermal pressure of a Fermi liquid of the form 
\cite{prakash_etal_1997}
\begin{equation}
    \label{eq:p_thermal}
    p_{th}^{\text{x}}(n_{\text{x}},T)
        = \frac{\pi^2}{6} n_{\text{x}} k T \frac{k T}{E_F^{\text{x}}}
\end{equation}
to the pressure at zero temperature.  Here, $n_{\text{x}}$ is the number
density of the relevant species, x=n, p, e and $\mu$ for neutrons, protons,
electrons and muons, respectively, $k$ is the Boltzmann constant and
$E_F^{\text{x}}$ is the Fermi energy of the given species. Note that, in our
case $kT/E_F^{\text{x}}\ll 1$ throughout most of the star (as our thermal
evolutions begin at $kT\approx 1$~MeV), whereas, e.g.,
\cite{lattimer_etal_1985} consider the thermal pressure when $kT \gtrsim
1$~MeV. As the electrons are relativistic, their Fermi energy is much higher
than that of protons or neutrons and as a result their contribution to the
thermal pressure is much lower in comparison (the situation is different when
the core is considered superfluid and the thermal pressure of protons and
neutrons is suppressed---the electron thermal pressure will dominate then).
Hence, we will only account for thermal pressure due to neutrons and protons.
The non-relativistic nucleon Fermi energy is given by $E_F^{\text{x}} =
p_{F\text{x}}^2 / 2m^*_{\text{x}}$ with $p_{F\text{x}} = \hbar k_F^{\text{x}}$
and $k_F^{\text{x}} = \left(3 \pi^2 n_{\text{x}}\right)^{1/3}$ and
$m^*_{\text{x}}$ being the Landau effective mass of the corresponding species
(for simplicity, taken to be constant in the following)
\cite{chamel_haensel_2006}. Since the EoS gives us information about the
composition of the neutron star core, we have all the information we need to
account for the thermal pressure due to neutrons and protons, and the total
pressure $p$ becomes
\begin{equation}
    p = p_0(\rho, x_\text{p})
        + \sum_{\text{x} = \text{n},\text{p}}
                p_{th}^{\text{x}}(n_{\text{x}}, T),
\end{equation}
where $p_0$ is the pressure of the zero temperature EoS
described in Section~\ref{ssec:eos} and $x_\text{p} = n_\text{p}/n_\text{b}$
is the proton fraction ($n_\text{b} = n_\text{n} + n_\text{p}$ is the baryon
number density). In order to see how the thermal pressure enters the
perturbation equations, we have to calculate the perturbed pressure. Since we
are assuming the composition to be frozen we have $\Delta x_\text{p} = 0$,
where $\Delta$ represents a Lagrangian perturbation. The perturbed pressure
then is
\begin{equation}
    \label{eq:delta_p}
    \Delta p
        = \left( \frac{\partial p_0}{\partial \rho} \right)_{x_\mathrm{p}}
            \Delta \rho
            + \left( \frac{\partial p_{th}^\text{n}}{\partial n_\mathrm{n}} \right)_T 
              \Delta n_\mathrm{n}
            + \left( \frac{\partial p_{th}^\text{n}}{\partial T} \right)_{n_\text{n}}
              \Delta T
            + \left( \frac{\partial p_{th}^\text{p}}{\partial n_\mathrm{p}} \right)_T 
              \Delta n_\mathrm{p}
            + \left( \frac{\partial p_{th}^\text{p}}{\partial T} \right)_{n_\text{p}}
              \Delta T. \\
\end{equation}

It is, of course, the case that the (cold) energy density $\rho$ is a function of the baryon density. Furthermore,  
keeping in mind that the composition is frozen and therefore protons and
neutrons are conserved separately, it follows~\cite{andersson_etal_2002} that
\begin{equation}
    \label{eq:same_ndensity_ratios}
    \frac{ \Delta n_\mathrm{n} }{ n_\mathrm{n} }
        = \frac{ \Delta n_\mathrm{p} }{ n_\mathrm{p} }
        = \frac{ \Delta n_\text{b}  }{ n_\text{b}   }
\end{equation}
since both species are comoving, too. We assume adiabatic oscillations,
$\Delta S = 0$, which allows us to replace the temperature perturbation, $\Delta
T$. The total entropy per baryon, $S$, is given by $S = S_\text{n} +
S_\text{p}$ with \cite{prakash_etal_1987}
\begin{equation}
    S_\text{x} = \frac{\pi^2}{2} \frac{T}{T_F^\text{x}},
\end{equation}
where the Fermi temperature, $T_F^\text{x} = E_F^\text{x} / k$, is a function of
the number density only, $T_F^\text{x} = T_F^\text{x} (n_\text{x})$. For
simplicity, we assume that the effective mass of both neutrons and protons
are identical, $m^*_\text{n} = m^*_\text{p} = m^*$. Adiabaticity then implies the simple relation
\begin{equation}
    \label{eq:dT_dn}
    \frac{\Delta T}{T} = \frac{2}{3} \frac{\Delta
    n_\text{x}}{n_\text{x}}\quad\text{ for }\text{x} = \text{n}, \text{p},
    \text{b}
\end{equation}
where we have used Equation~\eqref{eq:same_ndensity_ratios}.
From the thermal pressure given by Equation \eqref{eq:p_thermal}, it is easy to
show that
\begin{equation}
    \label{eq:dp_thermal}
    \left( \frac{\partial p_{th}^\text{x}}{\partial T} \right)_{n_\text{x}}
    = \frac{6 n_\text{x}}{T}
    \left( \frac{\partial p_{th}^\text{x}}{\partial n_\text{x}} \right)_T.
\end{equation}

Using Equations \eqref{eq:same_ndensity_ratios}, \eqref{eq:dT_dn} and
\eqref{eq:dp_thermal} we can eliminate $\Delta n_\mathrm{n}$, $\Delta
n_\mathrm{p}$ and $\Delta T$ from the perturbed pressure given in Equation
\eqref{eq:delta_p} in favour of $\Delta n_\text{b}$ and arrive at
\begin{equation}
    \Delta p
        = \left[
            \gamma p_0
            + 5 n_\mathrm{n} \left(
                    \frac{\partial p_{th}^{\text{n}}}{\partial n_\mathrm{n}}
                  \right)_T
            + 5 n_\mathrm{p} \left(
                    \frac{\partial p_{th}^{\text{p}}}{\partial n_\mathrm{p}}
                  \right)_T
          \right] \frac{\Delta n_\text{b}}{n_\text{b}},
\end{equation}
where we made use of the thermodynamic relation (for $\Delta T=0$)
\begin{equation}
    \Delta \rho = \frac{\rho + p}{n_\text{b}} \Delta n_\text{b}
\end{equation}
and the definition of $\gamma$ (see Equation \eqref{eq:def_gamma}). This
provides us with a straightforward way to incorporate the thermal pressure in
the perturbation problem. In our simulations, we account for the individual thermal pressures
of neutrons and protons in the core. In the crust, there are free
neutrons as well as ions present. For simplicity, we assume all baryons
contribute as a single Fermi liquid. This is a simplification which in essence implies a maximal thermal
component of the pressure.  This assumption will be relaxed in future work
using the results of, e.g., \cite{lattimer_etal_1985}.

\subsubsection{The Crust Elasticity}

It is not quite as straightforward to account for the elasticity of the star's
crust, at least not in general.  The approximation that the star is a perfect
fluid has to be abandoned as soon as the solid crust  forms, i.e. in the very
early stages of a neutron star's life. The elastic crust supports shear
stresses which by definition do not exist in a perfect fluid. To account for
such stresses, we have to introduce off-diagonal terms in the stress-energy
tensor. However, as we are assuming the background to be in a relaxed,
unstrained state, these alterations only appear in the perturbed stress-energy
tensor.

Following \cite{penner_etal_2011}, the shear strain tensor is given by
\begin{equation}
    \delta s_a^b
        = \frac{1}{2} \left(
            \perp^c_a \perp^{db} 
            - \frac{1}{3} \perp_a^b \perp^{cd}
          \right) \Delta g_{cd}.
\end{equation}
We will express  the metric perturbations in the well-known Regge-Wheeler
gauge \cite{regge_wheeler_1957}. This means that the components of the
Eulerian perturbations of  metric take the form
\begin{align}
    \label{eq:def_h}
    \delta g_{ab} = h_{ab} = -
    \begin{pmatrix}
        e^{\nu} r^l H_0(r)   & r^{l+1} \dot{H}_1(r)   &   0   &   0  \\
        r^{l+1} \dot{H}_1(r) & e^{\lambda} r^l H_2(r) &   0   &   0  \\
        0  &  0  &  r^{l+2} K(r)  &      0                           \\
        0  &  0  &       0        &  r^{l+2} \sin^2\theta K(r)
    \end{pmatrix}
    P_l(\cos\theta) e^{i \omega t},
\end{align}
where $P_l(\cos\theta)$ are the Legendre polynomials and dots denote time
derivatives. In the case of a
fluid source we have  $H_2=H_0$. The shear associated with the perturbations
enters the stress-energy tensor via the anisotropic stress tensor
\begin{equation}
    \delta \pi_a^b = - 2 \check{\mu} \delta s_a^b,
\end{equation}
where $\check{\mu}$ is the shear modulus. We have assumed a Hooke-like
relationship between the shear strain and stress. The total perturbed
stress-energy tensor then takes the form
\begin{equation}
    \label{eq:def_dT_tot}
    \delta T_{ab}^{\text{tot}} = \delta T_{ab} + \delta \pi_{ab},
\end{equation}
which leads to significant alterations of the perturbation equations compared
to the perfect fluid case. In order to deal with this, we define two new variables  related to the
traction. The radial traction, $T_1$, and the tangential traction, $T_2$,
are defined by
\begin{equation}
    \label{eq:def_traction}
    T_1 = \delta \pi_r^r\quad\text{and}\quad
    T_2 = \delta \pi_r^{\theta}.
\end{equation}
Both  variables vanish in the perfect fluid case (as $\check \mu \to 0$).
However, we will soon  see that they are crucial in the implementation of the
junction conditions at the crustal interfaces. The complete set of equations governing the perturbations in the elastic crust
are lengthy and do not provide deeper insight into the dynamics. We
therefore list them, together with their derivation, in
Appendix~\ref{sec:app_eq_crust_polar} and mention only the key features here.
The most notable difference between the two problems, key to including the
extra elastic dynamical degrees of freedom, is the fact that the the metric
perturbations $H_0$ and $H_2$ are no longer identical. Instead, we have
\begin{equation}
    H_2 = H_0 + 64 \pi \check{\mu} V \ , 
\end{equation}
where $V$ is related to the tangential displacement (see Equation
\eqref{eq:def_xi}).

\subsubsection{Phase-Transitions}

Finally, let us consider possible phase-transitions in the star, which may
come with a density discontinuity.  Having derived the perturbation equations
for the elastic crust, we need to connect the perturbations inside the crust
    to the perturbations for both the fluid core and the thin fluid ocean.
    This requires us to apply a set of interface conditions at each
    phase-transition. These conditions stem from the fact that the intrinsic
    curvature has to be continuous across the interfaces. This problem has
    already been analysed in detail \cite{finn_1990,penner_etal_2011}.  As our
    analysis is identical, we will simply state the relevant results here: At
    both crustal interfaces the continuity of the curvature imposes continuity
    of the perturbation variables $H_0$, $H_1$, $K$, $W$ and $T_2$. In addition to  the rather evident interfaces between the crust and the fluid,
we have to pay attention to other possible phase-transitions. As explained in
Section~\ref{ssec:eos}, the density profile inside a neutron star need not be
continuous;  phase-transitions at particular densities may cause density
discontinuities. Such discontinuities cause the perturbation
quantities to become discontinuous. However, it turns out that, as long as it
is only the density that is discontinuous, we arrive at the same jump
condition as for the crust-fluid interfaces (apart from the continuity of
$T_2$ which is trivially given as $T_2 = 0$ in the fluid anyway). Furthermore,
the Lagrangian pressure perturbation (the variable $X$, see
Section~\ref{sec:num_strategy}) is continuous across these interfaces.

\section{Numerical Strategy}
\label{sec:num_strategy}

The neutron star mode-problem requires the solution of a set of coupled first
order differential equations, and the identification of solutions that
represent purely outgoing gravitational waves at spatial infinity. The
technical issues associated with this problem have been discussed in detail
elsewhere, and the approach we take to solve it is relatively standard
\cite{thorne_campolattaro_1967,detweiler_lindblom_1985,finn_1986}.  Our
numerical strategy is conceptually the same as in several previous studies of
polar oscillations, the closest being Lin et~al.~\cite{lin_etal_2008} where a
three-layer star was considered.  The essential difference is that the
``middle'' layer in their study was considered superfluid whereas in our case
this layer forms the elastic crust.  Nevertheless, we still have to deal with
different sets of equations depending on whether a given layer is considered a
perfect fluid or  elastic  and the junction conditions required to connect the
layers.

We will outline the general idea for a specific case here and save a more
technical description for Appendix~\ref{sec:app_numerics}. We split the star
into several layers, so that each layer requires one unique set of equations
to be solved. Depending on the nature of each given layer, we have a certain
number (in our case four or six) of independent variables. We start
integrating the differential equations at one edge of the layer and continue
to the other end; repeating this procedure with linearly independent initial
vectors in order to provide us with a basis that can be used to express the
general solution to the problem.  After doing this for all different layers in
the star, we match the solutions at the interfaces. This leads to the need to
solve  a linear system for the ``amplitudes'' of the different solutions. The
boundary conditions at the centre and the surface of the star finally
determine one (up to amplitude) unique solution in the interior of the star.
At the surface, we match the solution to the Zerilli function which is solved
for outside the star and then calculate the amplitude of the ingoing wave
    amplitude at infinity, $A_\mathrm{in}$. Whenever this amplitude is zero,
    we have found a quasinormal mode of the star. 

The proposed approach has been used  successfully in a number of previous
studies \cite{detweiler_lindblom_1985,kokkotas_schutz_1992,comer_etal_1999},
but as soon as we consider low-frequency oscillations we  run into numerical
difficulties. Undesirable noise in the low frequency part of  the  spectrum
(higher order $g$-modes or thermal $g$-modes of old neutron stars as well as
some of the $i$-modes would lie in this regime) prevents us  from determining
any oscillation modes in this regime.  The  problem is of purely numerical
origin and  mainly stems from the algebraic equation used to calculate the
perturbation variable $V$ (cf. Equation~(A14) in
\cite{lindblom_detweiler_1983}):
\begin{equation}
    \label{eq:pfluid_V}
    \omega^2 \left( \rho + p \right) V
        = e^{\nu/2} X
            + \frac{1}{r} p' e^{\nu-\lambda/2} W
            - \frac{1}{2} \left( \rho + p \right) e^{\nu} H_0 ,
\end{equation}
where 
\begin{equation}
  X = - \frac{1}{r^l} e^{\nu/2} \Delta p . 
\end{equation}
A Taylor expansion (see \cite{detweiler_lindblom_1985})
reveals that all perturbation variables appearing in this equation, $H_0$,
$V$, $W$ and $X$, are of the same order of magnitude near the
origin. The same is true for the coefficients  on the right-hand side.
As we are dealing with low frequencies, $\omega M \approx 0.001$ (say), the left-hand
side of this algebraic relation forces the sum of the three terms on the
right-hand side to be many orders of magnitude smaller than the sum's
constituents, inevitably leading to numerical cancellation and hence
inaccurate results.

This problem has already been addressed by Finn \cite{finn_1986}, who proposed
a different set of equations for the study of low-frequency modes. His
formulation makes use of the Eulerian pressure perturbation rather than the
Lagrangian one. Such an  implementation improves the accuracy of our results
as the effect of the cancellation is not as severe; however, it does not
fully satisfy our demands.  Instead, we derive a new set of equations where
the independent variables are a subset of the `fundamental' variables, namely
$H_1$, $K$, $V$ and $W$.  This formulation does not involve an algebraic
equation like Equation \eqref{eq:pfluid_V} and thus the cancellation is
avoided.  The complete set of perturbation equations we use is given in
Appendix~\ref{sec:app_eq_low_freq}. The price we pay for eliminating the
cancellation is the appearance of the derivative $\rho'$ in the perturbation
equations. Since this quantity is not known during the integration of the TOV
equations, we have to calculate it numerically on an uneven grid; even worse,
it may not be well-defined everywhere since $\rho(r)$ is not necessarily
differentiable (density discontinuities may occur inside the star). We use the
method proposed by Sundqvist \& Veronis \cite{sundqvist_veronis_1970} to
tackle the former problem. In order to overcome the latter, we only use this
set of equations in a region near the origin up to a certain radius $R_V$, at
which point  the magnitude of the variables have changed  sufficiently that
Equation \eqref{eq:pfluid_V} can be applied without significant loss of
accuracy and then switch back to the perfect fluid equations from
\cite{detweiler_lindblom_1985}. This encourages a rather large value for
$R_V$. However, when choosing $R_V$, we also have to ensure that there are no
discontinuities in $\rho$ inside the interval $[0,R_V]$ and furthermore, the
calculation of $\rho'$ suffers from (small but nevertheless present) numerical
errors, which favours a small value for $R_V$. In the end, the `optimal' value
for $R_V$ requires trial and error. In our calculations we find that $R_V
    \approx 5\unit{km}$ is a good choice as it allows us to reliably extract
    mode frequencies down to $\approx 18\unit{Hz}$ (the actual lower limit
    slightly depends on the problem under consideration, see Section
    \ref{sec:results}). For different values of $R_V$, the numerical issues
    become more severe and in the end the low frequency spectrum is
    contaminated by noise up to higher frequencies. Further work is needed
    on this problem.

\section{Results}
\label{sec:results}

Since the spectrum of a relativistic star is  rich, and could be difficult to
untangle, we will build our understanding by taking the key bits of physics
into account step by step. This will gradually enrich the spectrum with more
and more classes of oscillation modes. The natural steps are: (i) taking only
composition gradients into account, (ii) switching on temperature using the
thermal evolution and (iii) accounting for the crystallisation of the elastic
crust. The density jumps, leading to $i$-modes, are built-in to the EoS, so we
cannot easily switch them on or off. Hence, these  modes will always be
present in the star's spectrum.

Let us briefly comment on the actual implementation of the density
discontinuities. By far the easiest way of ``implementing'' them is to simply
have them present in the tabulated EoS; for each discontinuity there are two
adjacent rows in the given EoS table with nearly (but not actually) identical
values for the pressure but the density changes considerably. A TOV solver
with adaptive mesh refinement will then lower the step size around these
densities and the background model will have large gradients, $\rho'$, at
these densities. However, the phase-transitions do not appear as proper
discontinuities here. It turns out that this is enough to find the interface
modes in the spectrum. We also take a different approach to ensure that every
phase-transition is reflected by an actual jump in density in our background
model.  For this, we modified the tabulated EoS by replacing the two pressure
values for each phase-transition in the two adjacent rows by their average
value.  We also adjusted our TOV solver to ensure that two grid points are
placed at each phase-transition; both with the same pressure but with the
highest and lowest density of the discontinuitiy interval, respectively. When
we compare these two implementations, we find that the frequencies of the
interface modes with high frequencies are hardly changed while the interface
modes with lower frequencies experience a serious shift.

In our calculations, we have to opt for one of these approaches. While
the approach with actual discontinuities in density in the background model
has the advantage of having a somewhat nicer numerical solution (there won't
be sharp spikes in the density gradient), it is certainly the case that a
phase-transition in an actual neutron star occurs over a small but non-zero
distance whereby the ``simple implementation'' is physically favoured. In our
results, we opt for using the EoS with actual discontinuities but we point out
that a switch to the other case can be done quickly. Jog \& Smith
\cite{jog_smith_1982} argued that a phase transition between two layers occurs
over a very narrow pressure range $\Delta p$. Typically, this pressure range
$\Delta p$ is about four orders of magnitude smaller than the pressure $p$ at
which it occurs: $\Delta p \approx 10^{-4} p$. Hence, the approximation of
implementing sharp density jumps is quite a good representation of the nuclear
composition in the crust.

As in earlier work \cite{detweiler_lindblom_1985,andersson_etal_1995}, and
since our focus is on oscillations that are slowly damped by
gravitational-wave emission, we will not actually calculate the complex
frequencies of the quasinormal modes; we ignore the damping of the mode.
Instead, we construct the asymptotic amplitude $A_{\text{in}}$ for real-valued
frequencies and locate the quasinormal modes by approximating  the zeros of
this function (essentially resonances in the problem). To visualise the
spectrum, we plot the logarithm of the incoming amplitude,
$\log|A_\text{in}|$, as a function of the frequency (as in
Figures~\ref{fig:i-modes-composition}, \ref{fig:g-modes},
\ref{fig:p-evolution} and \ref{fig:s-modes}); the
logarithm turns the zeros into much more visible spikes and an eigenfrequency
can be found where this function tends to $-\infty$. The damping
times can also be estimated using this procedure \cite{andersson_etal_2002}
but we will not do so here. In fact, the damping times of (most of) the modes
under consideration are so long, i.e. the imaginary part of $\omega$ is many
orders of magnitude smaller than its real part, that a reliable calculation of
the damping times would be impossible due to the finite machine precision.

We also comment here on the stability of our numerical procedure. As explained
earlier,  the calculation of the ingoing wave amplitude is spoilt by noise in
the low frequency regime and, in practice, the lower limit for reliably
extracting frequencies lies at about $18\unit{Hz}$. In our simulations, we
find that this lower limit gets shifted down to approximately $10\unit{Hz}$
when we include composition, i.e.  when we use $\gamma$ instead of $\gamma_0$
in the equations (see next subsection). This gives us a further hint to the
origin of the instability in the equations. However, this is not the only
crucial point since the noisy behaviour is almost non-existent for polytropic
equations of state (which we  implemented as a test of the numerics).

\subsection{Composition Gradient}

\begin{figure}[ht]
    \centering
    \includegraphics[clip=true,width=0.6\textwidth]{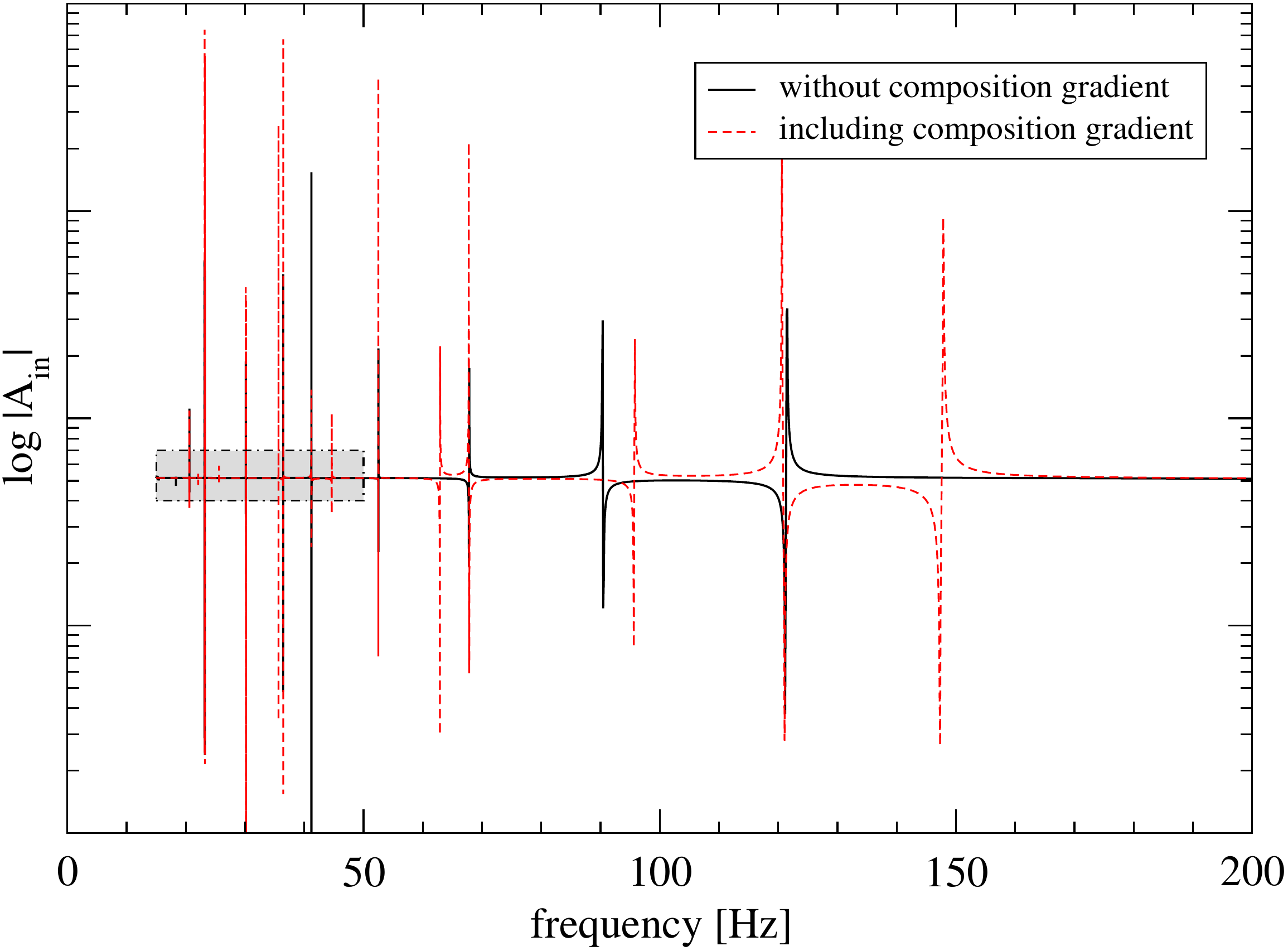}

    \vspace{1cm}

    \includegraphics[clip=true,width=0.6\textwidth]{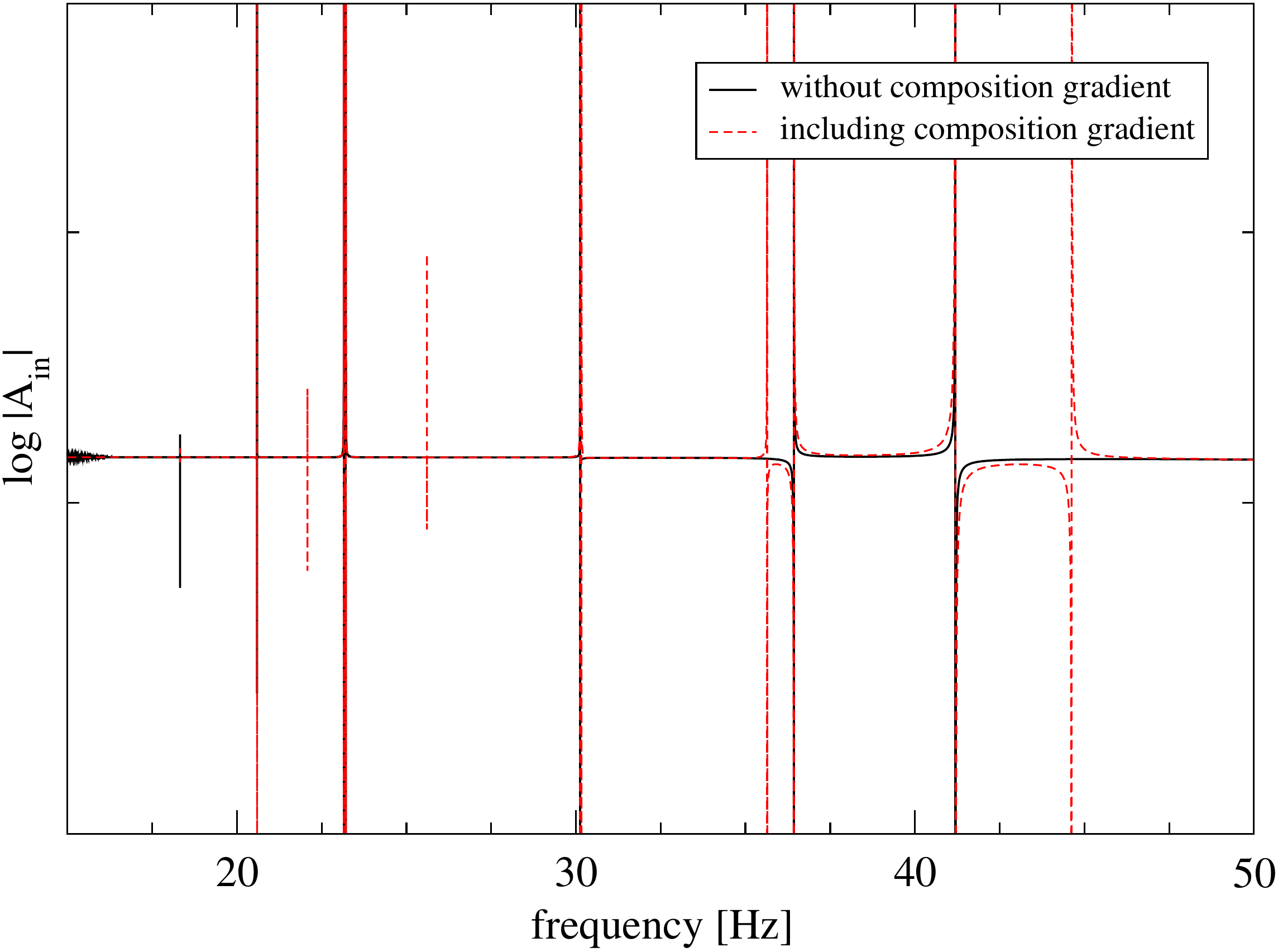}
    \caption{The low frequency spectrum  of our neutron star at zero
    temperature and without a solid crust. The upper panel shows the spectrum
    up to $200\unit{Hz}$ whereas the lower panel shows a
    magnification of the grey shaded area. The solid line shows the spectrum
    of the pure perfect fluid star, while the dashed line includes composition
    gradients. All spikes in the solid spectrum are interface modes as there
    is no composition gradient present; for each of these modes, there is also
    a mode present in the spectrum of the stratified star (with the exception
    of the mode at $90.4\unit{Hz}$; see text). The ``new'' modes in the dashed
    spectrum are composition $g$-modes.}
    \label{fig:i-modes-composition}
\end{figure}

The first step beyond barotropic models involves including the composition
gradient. Then we expect to find composition $g$-modes in the spectrum as well
as $i$-modes due to the density discontinuities. The high
frequency domain will be populated by $f$-modes and $p$-modes but we will not
discuss this part of the spectrum much as it is well understood
\cite{cowling_1941,detweiler_lindblom_1985}. Instead, we show the
low frequency domain up to $200\unit{Hz}$ (as well as a zoom in at the lower
end of the frequency window) for our neutron
star model with zero temperature and without solid crust in
Figure~\ref{fig:i-modes-composition}.
This frequency range contains all low frequency modes in the model (above the
noise cut-off).  For comparison we show the spectrum for both the unstratified
star (solid line) and the corresponding stratified star (dashed line). All
low-frequency modes in an unstratified star are $i$-modes due to density
discontinuities and there is precisely one interface mode associated with each
density discontinuity. We order them by frequency and label them as $i_n$ ($n
\ge 1$), where $i_1$ is the interface mode with the highest frequency.

\begin{figure}[ht]
    \centering
    \includegraphics[clip=true,width=0.6\textwidth]{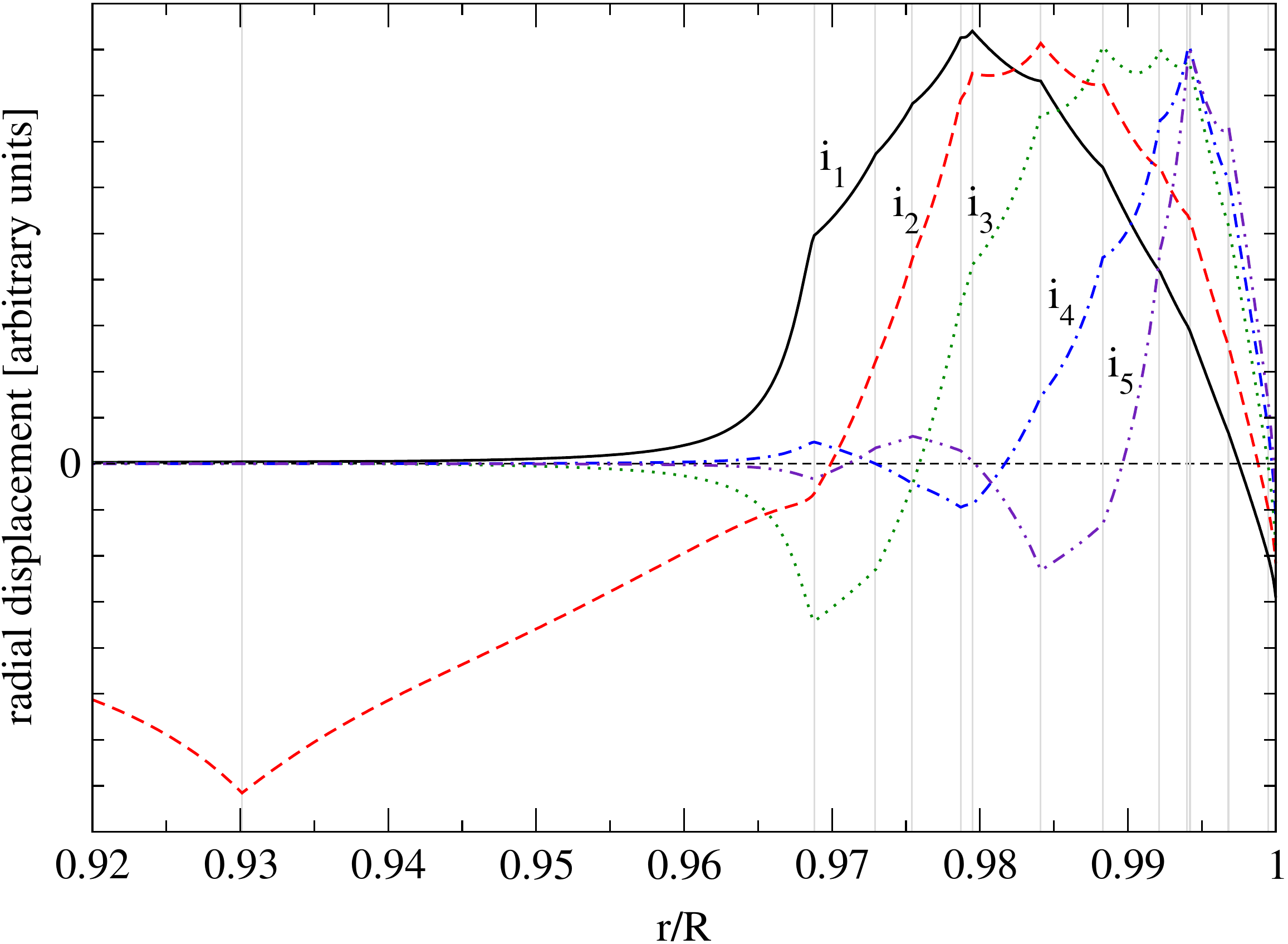}
    \caption{The radial displacement, $\xi^r$, for five $i$-modes of the
    unstratified star. The vertical, grey lines show the location of the
    different density discontinuities within the star; the interface at $r/R
    \approx 93\,\%$ is the crust-core interface. The amplitudes of the
    different modes are scaled so that they are of comparable size. The kinks
    at the phase-transitions are clearly visible and the displacement nearly
    vanishes in the core (with the exception of the $i_2$-mode which is
    associated with the crust-core transition). The eigenfunctions are largely
    unaffected if stratification is switched on (again with the exception of
    the $i_2$-mode which turns into a $g$-mode, see text).}
    \label{fig:i-modes_ef}
\end{figure}

It turns out that the association with interface modes and specific density
discontinuities is not unambiguous. The general rule that the radial
displacement of an interface mode has its peak amplitude exactly at the
corresponding phase-transition does not hold in our case. The eigenfunctions
do have their maximum at some phase-transition but we find that several
interface modes would belong to the same phase-transition according to this
procedure. Likewise, there are discontinuities at which none of the $i$-modes
have a maximum.
This is the result of the presence of several discontinuities in close
vicinity which affect each other. We show the eigenfunctions of five
interface modes in Figure~\ref{fig:i-modes_ef}, where the kinks at the
phase-transitions are clearly visible. The modes $i_4$ and $i_5$ have their
strongest peak at the same transition. However, the $i_1$ mode has its highest
amplitude at a transition where no other interface mode (also the ones which
are not shown) has its strongest
peak. This suggests that this mode is associated with the transition
${}^{80}\text{Zn} \rightarrow {}^{78}\text{Ni}$. The $i_2$-mode apparently has
its most prominent peak much deeper in the star and is associated with the
artificial discontinuity at the crust-core interface due to the matching of
the EoS.

Finn \cite{finn_1987} provides a simple formula to give an estimate for the
frequency of an $i$-mode; his calculations show that---in an idealised
situation---the frequency of the interface mode is proportional to the
relative jump in density, $\Delta\rho / \rho_+$, and the distance, $\Delta r$,
of the associated phase-transition from the surface of the star.  Even though
our results qualitatively agree with this estimate, the formula is,
unfortunately, not accurate enough to make the exact assocation. Exemplary for
the qualitative agreement, we inspect the mode $i_1$, which has the highest
frequency amongst the interfaces modes in the star. It is due to the
transition, which, when compared to the other phase-transition, has the
largest relative jump in density of $\Delta\rho / \rho_+ \approx 4.3\,\%$ and
is located rather deep in the star.

All interface modes (except for the one at $90.4\unit{Hz}$; see discussion
below) are still present in the stratified star. However, we find a set of new
modes appearing in the spectrum; these are the composition $g$-modes due to
the stratification of the core (composition variations in the crust
distinguish different layers and so only lead to interface modes associated
with transitions from layer to layer). $g$-modes and $i$-modes are easy to
distinguish if one examines their eigenfunctions.  While the $g$-modes are
mainly confined to the core and their radial displacement is a smooth function
of the radius, the radial displacement of the $i$-modes nearly vanishes in the
core and exhibit kinks at the phase-transitions (cf.
Figure~\ref{fig:i-modes_ef}). In contrast to the $i$-modes, the set of
$g$-modes is infinite and we label them as $g_n$, where $g_0$ is the $g$-mode
with the highest frequency and $n$ will correspond to the number of nodes of
the radial displacement in the core.

\begin{table}[ht]
  \caption{Frequencies of the $i$-modes and composition $g$-modes in the
  stratified, cold star.}
  \centering
  \begin{tabular}{c c | c c}
    \hline
    Frequency & Mode  & Frequency & Mode \\
      {[}Hz]  &       &   [Hz]    &      \\ [0.5ex]
    \hline
        147.3 & $g_0$ & 30.2 & $g_5$    \\
        121.1 & $i_1$ & 30.1 & $i_6$    \\
         95.6 & $g_1$ & 25.6 & $g_6$    \\
         67.8 & $i_2$ & 23.2 & $i_7$    \\
         62.8 & $g_2$ & 23.1 & $i_8$    \\
         52.5 & $i_3$ & 22.1 & $g_7$    \\
         44.6 & $g_3$ & 20.6 & $i_9$    \\
         41.2 & $i_4$ & 19.6 & $g_8$    \\
         36.4 & $i_5$ & 18.3 & $i_{10}$ \\
         35.6 & $g_4$ &      &          \\

    \hline
  \end{tabular}
  \label{tab:i-modes}
\end{table}

Starting from the surface of the star, the 13th jump in density, which is due
to the manual matching of the EoS at the crust-core interface, deserves a few
more comments since no associated interface mode can be found in the
stratified star. When the star is barotropic (not stratified), we find an
$i$-mode at $90.4\unit{Hz}$ associated with this density discontinuity. As this
discontinuity lies within (or at least at the edge of) the region where we
account for stratification and an interface mode can be understood as a
special kind of composition $g$-mode (a sudden change in density indicates a
phase-transition and as a result, a perturbed fluid element crossing this
interface will experience buoyancy due to the different composition), this
$i$-mode loses its character and turns into the lowest order $g$-mode when
the star is stratified. To verify this, we ran a series of separate
simulations in which we slowly increased the composition gradient via the
adiabatic index $\gamma$ from ``no stratification'' up to ``full
stratification'' (as given in the EoS). The interface mode accordingly
increased in frequency and its eigenfunction changed,  turning
into the $g_0$ mode of the stratified star. This example illustrates the close
relationship between these two classes of modes.

We list the low frequency modes of the stratified star in
Table~\ref{tab:i-modes}.  The spectrum of the barotropic star can be easily
extracted from this table by the following procedure: Remove all $g$-modes
from this table and add one $i$-mode at $90.4\unit{Hz}$. The frequencies of
the $i$-modes are barely affected by stratification (they vary only by a few
tenths of a Hertz).

\subsection{Finite Temperature}

\begin{figure}[t]
    \centering
    \includegraphics[clip=true,width=0.6\textwidth]{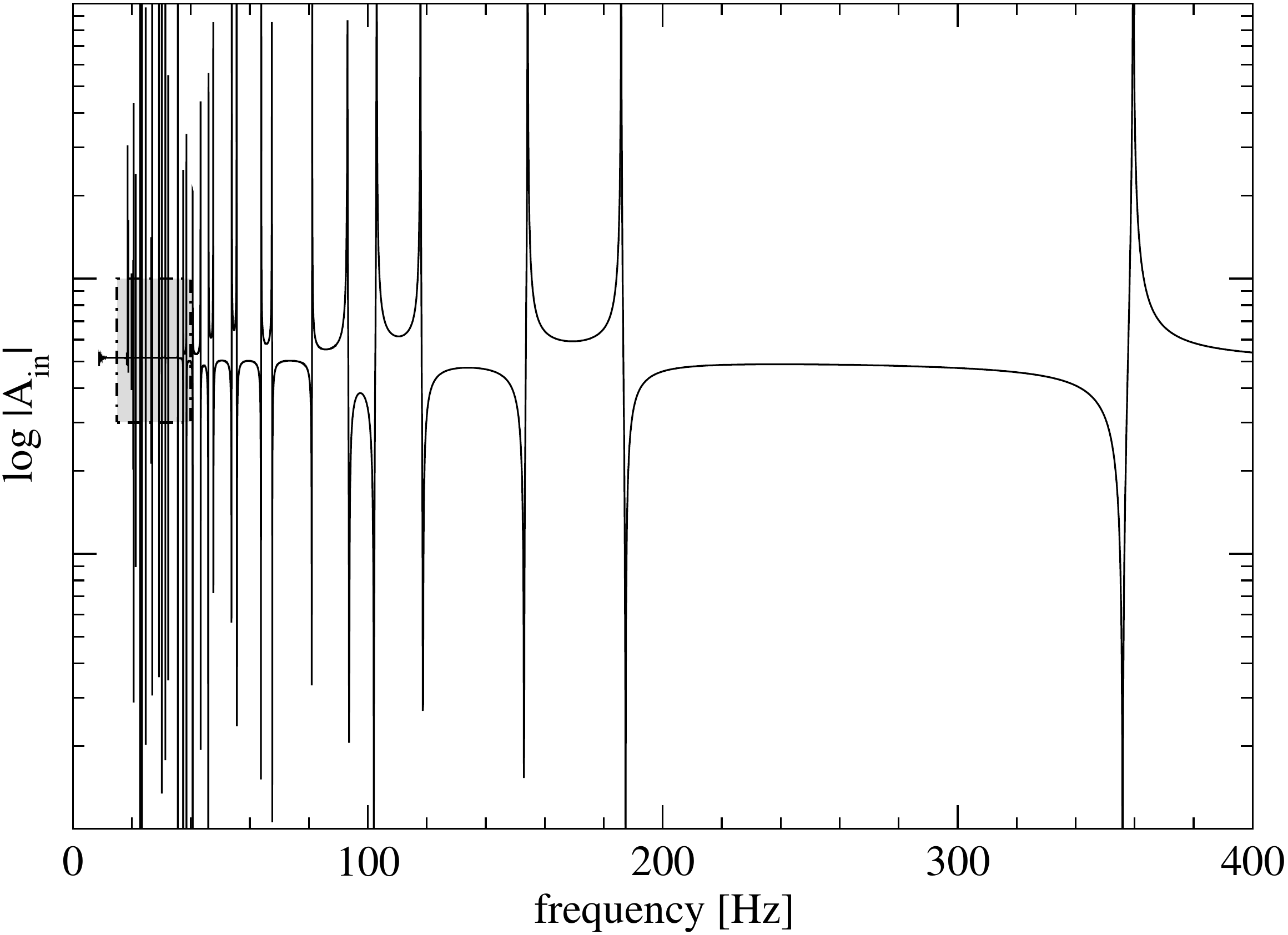}

    \vspace{1cm}

    \includegraphics[clip=true,width=0.6\textwidth]{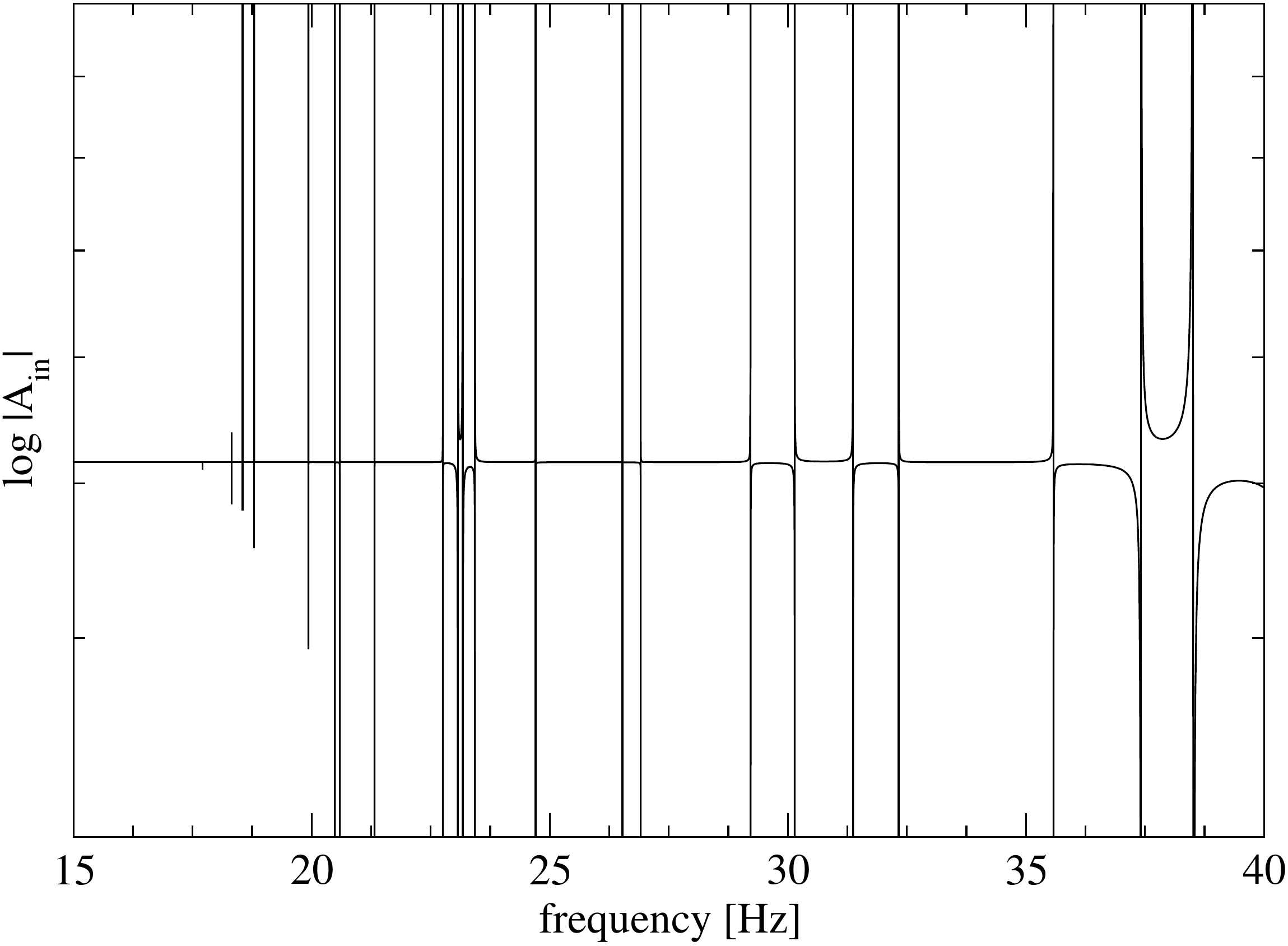}
    \caption{The low frequency spectrum when the star is 3 seconds old;
    including thermal pressure but without solid crust. The
    lower panel shows a magnification of the grey shaded area in the upper
    panel.}
    \label{fig:g-modes}
\end{figure}

The next step is to move away from the assumption that the neutron star is
cold. As we have already described, we account for the thermal pressure by
adding it to the static pressure of the cold EoS used to
determine the background. This leads to the results shown in
Figure~\ref{fig:g-modes}, which shows the low frequency spectrum of an
isothermal neutron star with temperature $Te^{\nu} = 10^{10}\unit{K}$ (our
initial configuration). In comparison to the result at zero temperature
(cf.~Figure~\ref{fig:i-modes-composition}), we find a vastly enriched
spectrum.  A careful investigation reveals that all composition $g$-modes
found in the cold, stratified star are still present in the hot star at only
marginally altered frequencies (the difference is generally less than 0.1\%,
only the high freqency $g$-modes are shifted by up to 4\%); likewise their
eigenfunctions are unaltered. This is as expected since the composition
$g$-modes in our model originate from the composition gradient in the core.
The thermal pressure, however, is negligible in the core and hence does not
affect the $g$-modes. 

For the $i$-modes the situation is somewhat different. We find all interface
modes in the hot star, too; however, while the low frequency interface modes
maintain their frequency throughout the thermal evolution of the star, four
interface modes with high frequencies ($i_1$, $i_2$, $i_4$ and $i_5$) have
their oscillation period decreased. We are unable to provide a reason as to
why this behaviour is observed for precisely these interface modes but not
others.

\begin{figure}[ht]
    \centering
    \includegraphics[clip=true,width=0.6\textwidth]{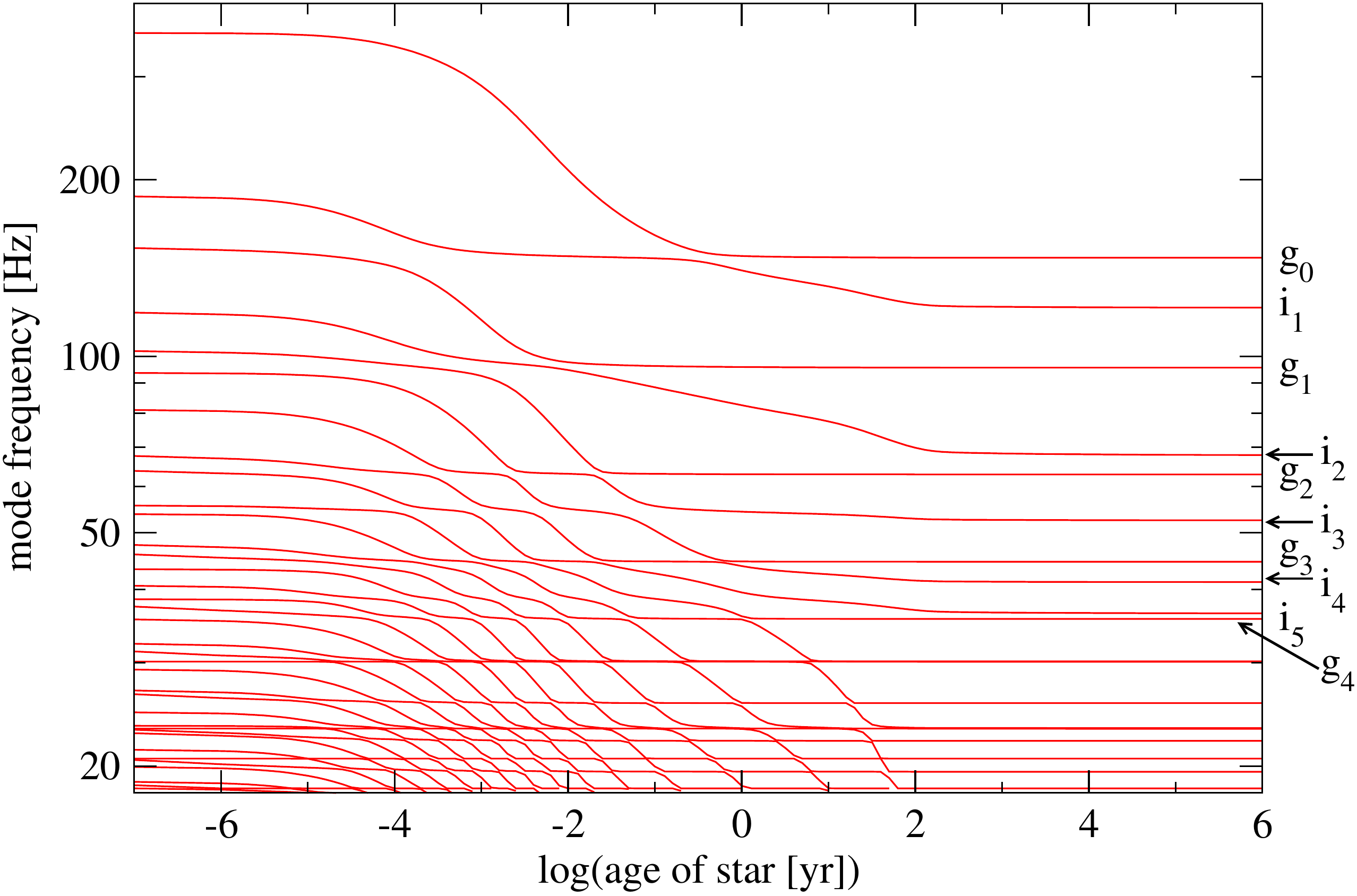}
    \caption{The low-frequency spectrum as the star cools. The
    thermal $g$-modes quickly decrease in frequency and fall below
    $18\unit{Hz}$ after about 100 years. Four of the high frequency interface
    modes are affected by the thermal pressure ($i_1$, $i_2$, $i_4$ and
    $i_5$). All modes clearly exhibit avoided crossings.}
    \label{fig:g-evolution}
\end{figure}

In order to track the evolution of the spectrum as the star ages, we produced
a time evolution of the low frequency modes in the following way: We start
with the isothermal star at $Te^{\nu} = 10^{10}\unit{K}$, for which the
spectrum is that shown in Figure~\ref{fig:g-modes}, and extract temperature
profiles at 120 time steps  uniformly distributed on a logarithmic timescale.
After extracting the mode frequencies for each of these temperature profiles,
we are able to trace the evolution of the oscillation modes over time. This
leads to the results shown in Figure~\ref{fig:g-evolution}.  All modes exhibit
avoided crossings which are easily visible in the high frequency part of the
graph; they are also present in the low frequency part, where a higher
resolution is necessary to resolve the different modes.

We now find a set of new modes spread over the entire spectrum;
these are the thermal $g$-modes. Their frequency decreases as the star cools
and after about 100 years nearly all the thermal $g$-modes have frequencies of
$18\unit{Hz}$ or less. This is when the temperature has decreased so far that
the thermal pressure is almost negligible and does not affect the frequencies
of the $g$- and $i$-modes anymore.

\begin{figure}[ht]
    \centering
    \includegraphics[clip=true,width=0.6\textwidth]{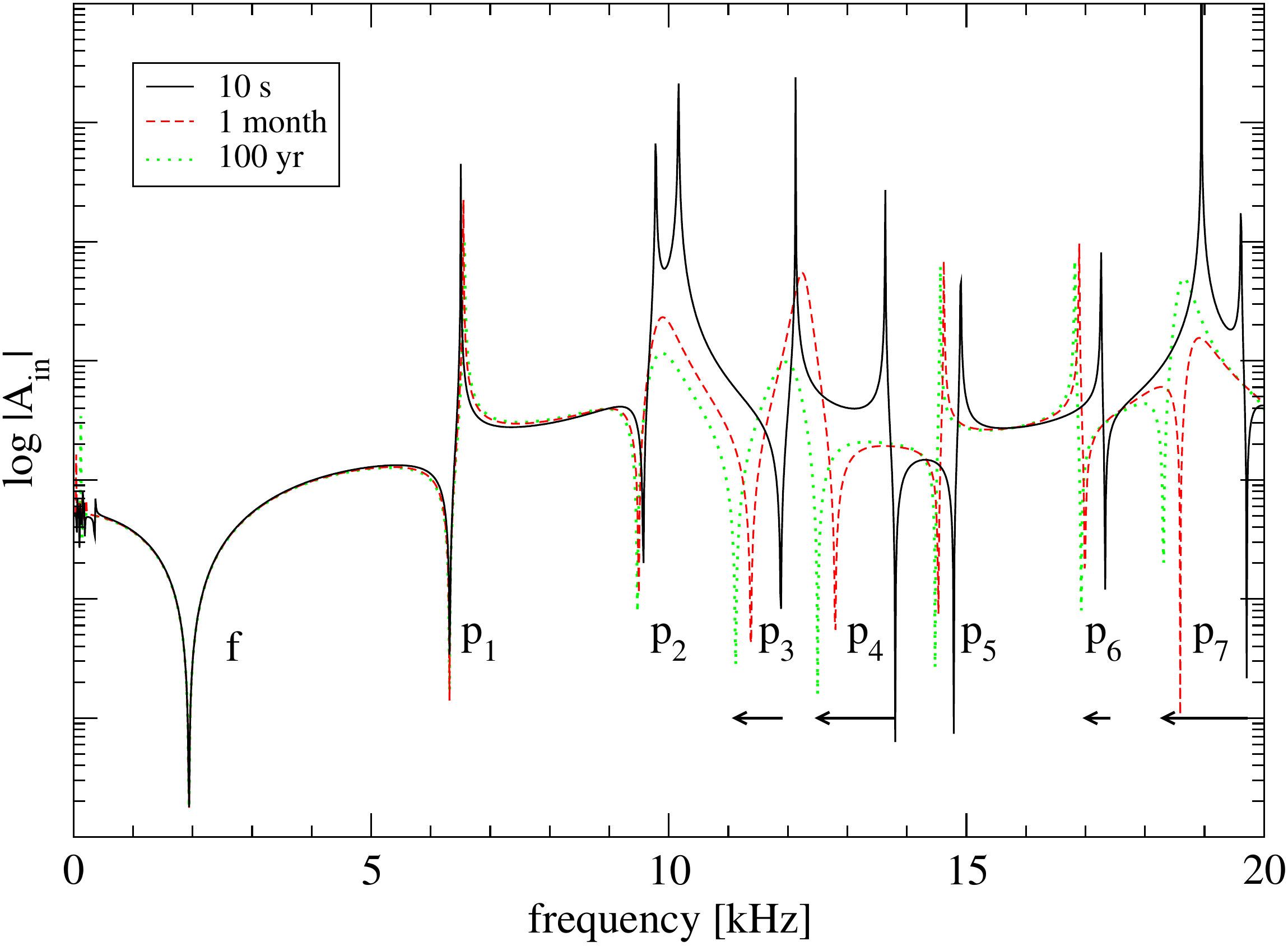}
    \caption{The stellar spectrum up to $20\unit{kHz}$ for three different
    temperature profiles. Clearly visible are the $f$-mode and first 7
    $p$-modes. Where the frequency of a $p$-mode varies visibly over time, an
    arrow indicates this change.}
    \label{fig:p-evolution}
\end{figure}

For completeness, we considered the impact of  thermal pressure on the
fundamental and pressure modes. As an illustration of the results, we show the
star's spectrum after 10 seconds, one month and 100 years in
Figure~\ref{fig:p-evolution}. Since the $f$-mode is mainly due to the surface
of the star and depends on the average density, it does not change (especially
since we do not account for thermal pressure in the background model). The
$p$-modes, however, should be affected since the thermal pressure contributes
to the restoring force acting on sound waves. In our simulations we observe a
slight increase in frequency by up to roughly 10\,\%, but only for $p$-modes of
higher order and only in young, hot neutron stars. As is apparent from
Figure~\ref{fig:p-evolution}, after just one month the frequencies of the
$p$-modes are only slightly affected by the thermal pressure.

\subsection{The Crust Elasticity}

\begin{figure}[ht]
    \centering
    \includegraphics[clip=true,width=0.6\textwidth]{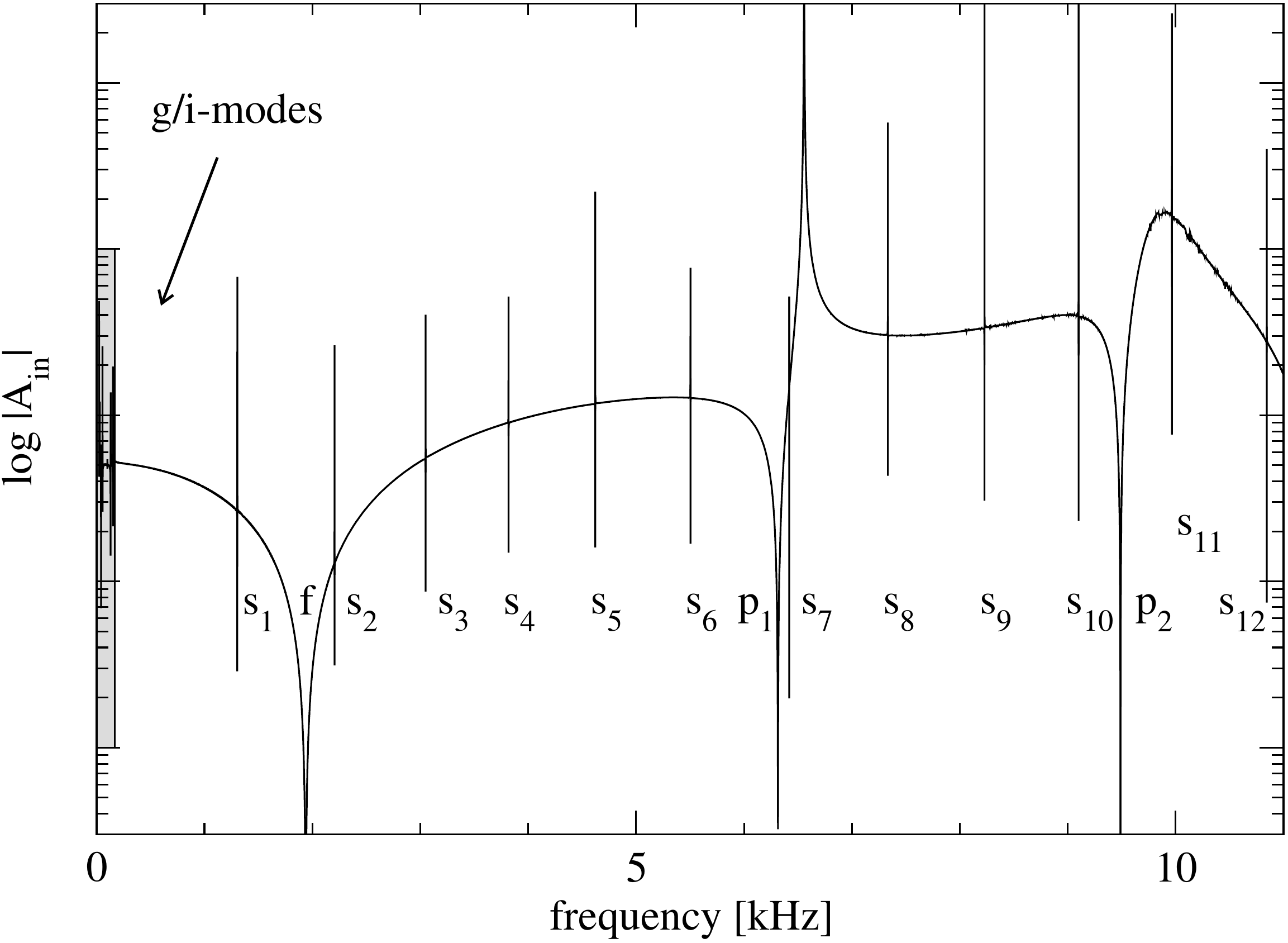}

    \vspace{1cm}

    \includegraphics[clip=true,width=0.6\textwidth]{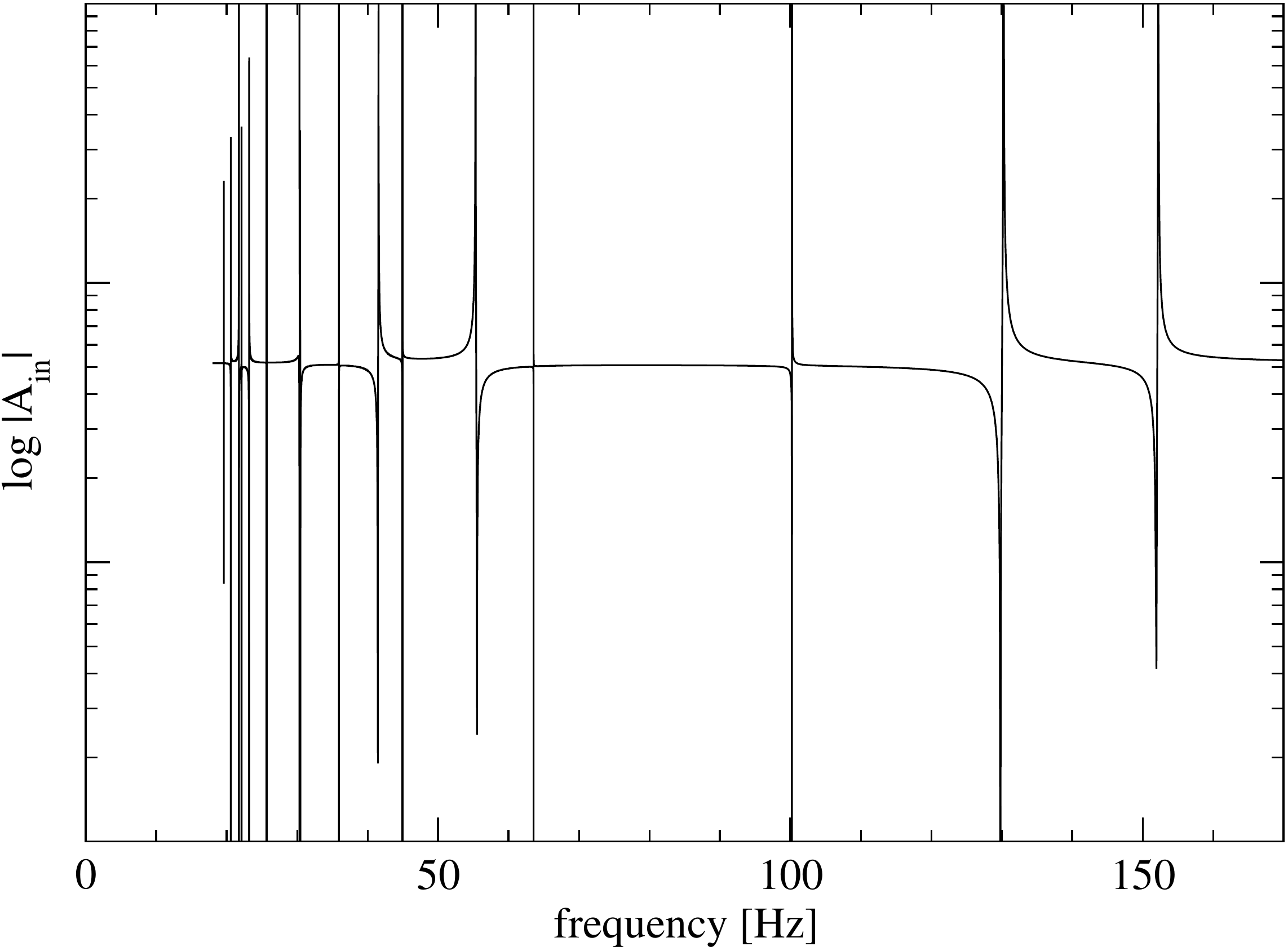}
    \caption{The spectrum of our model star including thermal pressure and an
    elastic crust, after hundred years of the star's life. The upper panel
    shows the high frequency domain, while the lower panel zooms into the grey
    shaded area at low frequencies to resolve the $g$- and $i$-modes.}
    \label{fig:s-modes}
\end{figure}

Finally, we add the elastic crust to the model. As explained in Section
\ref{ssec:crust_formation}, we define the region in which the crust is solid
via the condition $\Gamma > 173$ for the Coulomb coupling parameter. As in the
previous Section, we start from an isothermal star with $Te^{\nu} =
10^{10}\unit{K}$. At this point, the crust is above its melting temperature
and hence liquid; after approximately 1.1 days, the crust starts to
crystallise (cf. Figure~\ref{fig:crust_formation}). In
Figure~\ref{fig:s-modes} we show the spectrum up to $8\unit{kHz}$ for the
neutron star at an age of 100 years, including the thermal pressure and
composition gradients. The lower panel in Figure~\ref{fig:s-modes} provides a
zoom-in at low frequencies. The two most prominent oscillation modes at
$1.938\unit{kHz}$ and $6.315\unit{kHz}$ are the $f$-mode and the first
$p$-mode, respectively; see also Figure~\ref{fig:p-evolution}. As expected,
they are unaffected by the presence of the solid crust. 

\begin{figure}[ht]
    \centering
    \includegraphics[clip=true,width=0.6\textwidth]{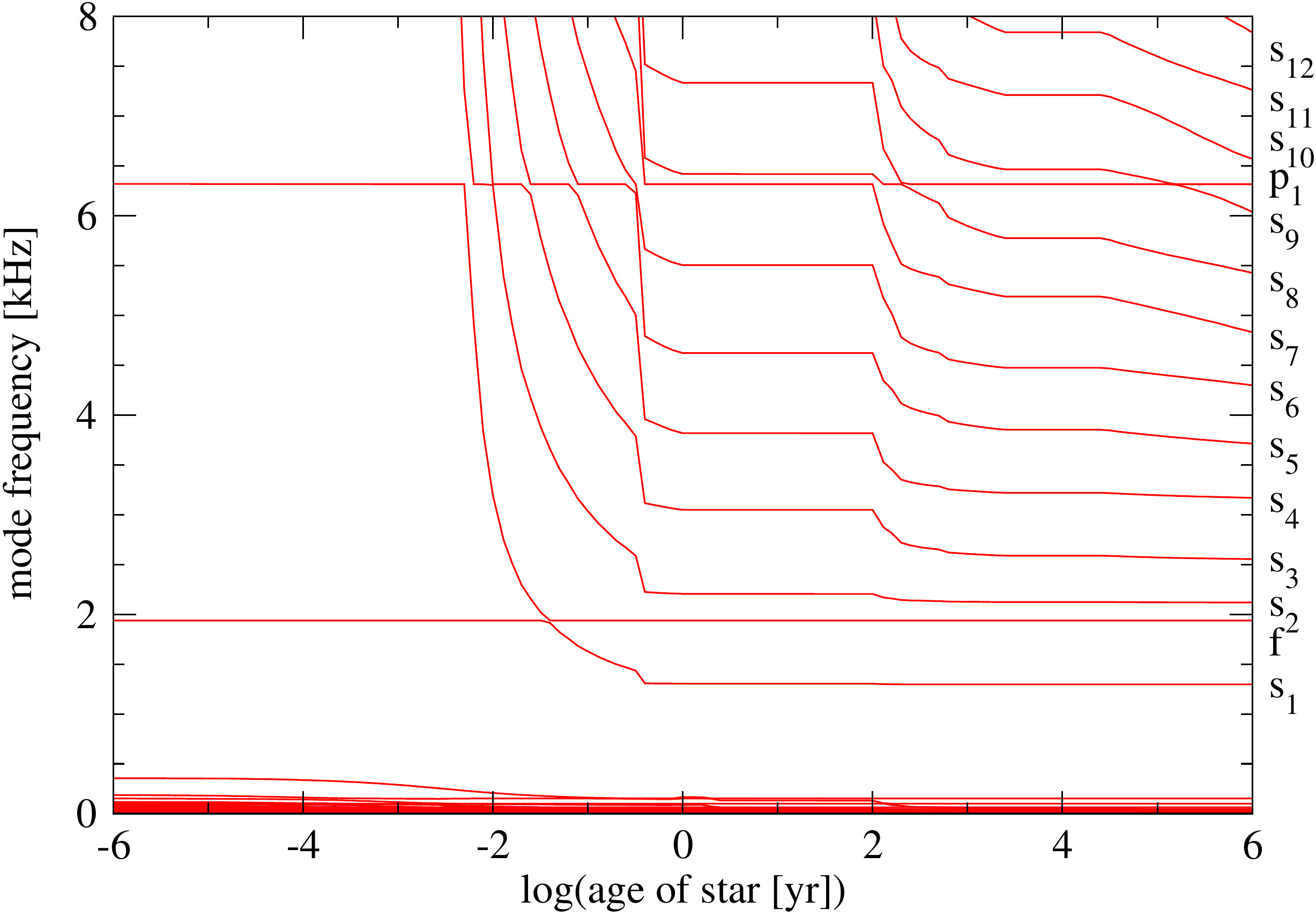}

    \vspace{1cm}

    \includegraphics[clip=true,width=0.6\textwidth]{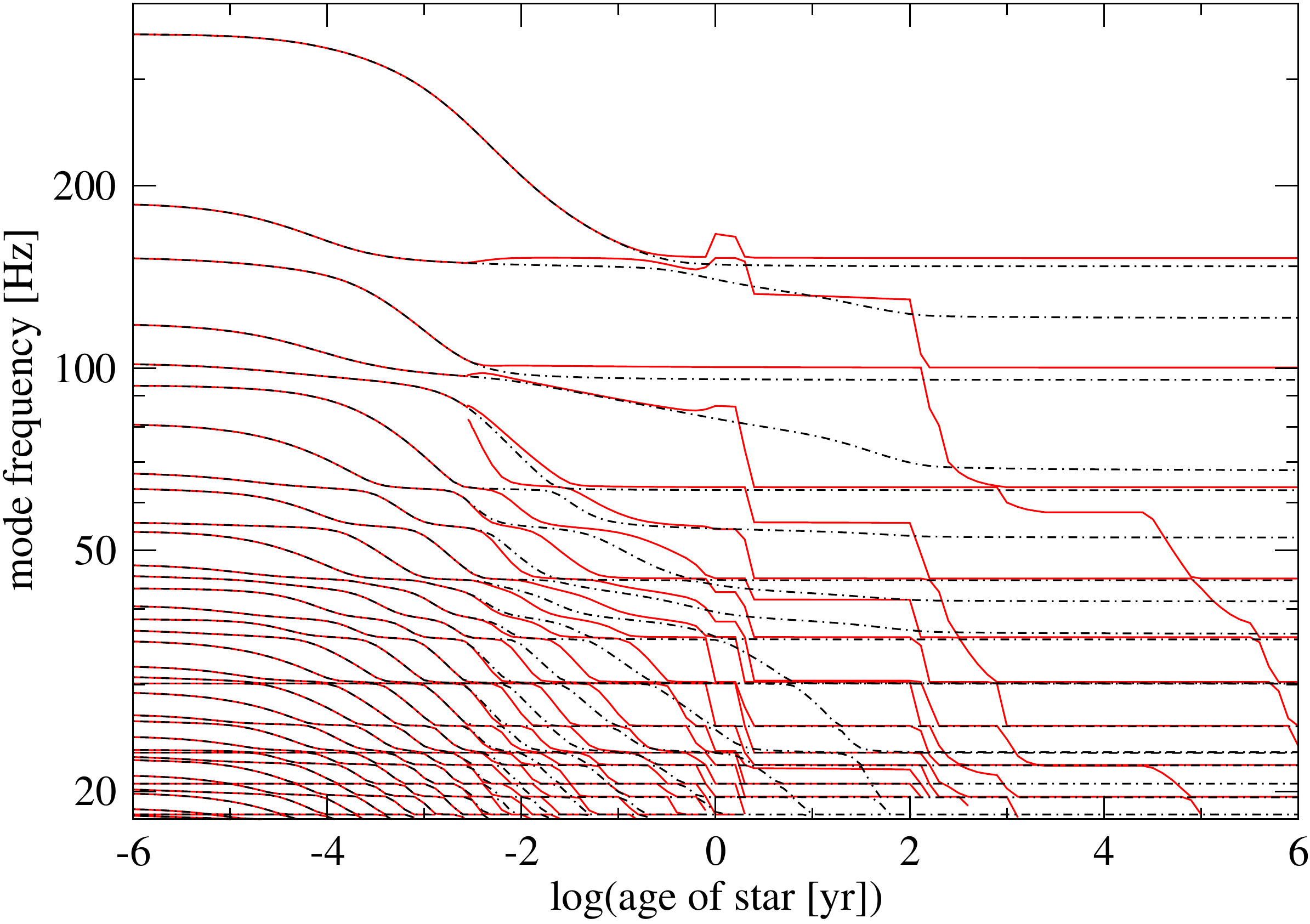}
    \caption{This figure shows how the mode frequencies of the star with
    elastic crust change as the star cools.
    \emph{Upper panel:} Same as Figure~\ref{fig:g-evolution} but this
    time for the high frequency part of the spectrum and the model star
    accounts for the elastic crust. The two modes at constant frequencies,
    $1.938\unit{kHz}$ and $6.315\unit{kHz}$, are the $f$-mode and first
    $p$-mode, respectively, whose frequencies remain unaffected. The shear
    modes appear at the top end of the spectrum after about 1.1 days when the
    crust starts to crystallise. At the bottom of the graph the low frequency
    $g$-modes and interface modes are visible; see lower panel.
    \emph{Lower panel:} A zoom-in at low frequencies in order to resolve the
    numerous $g$-modes and interface modes. We include the evolution of the
    low frequency spectrum of the star without elastic crust (dash-dotted
    lines) for comparison. As the crust of our model star is fluid during its
    very early stages, the evolutions coincide for about 1.1 days until the
    crust starts to crystallise. Then the evolution of the low frequency modes
    is affected by the solid crust, leading to somewhat higher mode
    frequencies.
    }
    \label{fig:s-evolution}
\end{figure}

The low frequency domain again shows evidence for a large set of oscillation
modes. However, only very few modes have maintained their frequency compared
to the perfect fluid star. This is not surprising as we have changed the
physics of the crust considerably and all the interface modes stem from
phase-transitions within the crust. We show the evolution of the low frequency
spectrum of the star with elastic crust in the lower panel of
Figure~\ref{fig:s-evolution}; for comparison we include the evolution of the
star without elastic crust (the dash-dotted lines). Since the crust
crystallises only after about 1.1 days, both evolutions coincide in the very
early stages of the star's life. As soon as the crust exhibits elastic parts,
the spectra start to diverge from each other. The modes at higher frequencies
tend to slightly increase in frequency, whereas the modes at lower frequencies
are shifted to lower frequencies.

The threshold between these to different characteristics is the mode at
$86.2\unit{Hz}$ appearing after 1 day (at the time $\log(t/\unit{yr}) = -2.56$
in Figure~\ref{fig:s-evolution}); as soon as the crust solidifies, this mode
gets shifted up to $86.7\unit{Hz}$ and a new mode appears in close vincinity
at $82.3\unit{Hz}$ which cannot be directly linked
to any other oscillation mode present in the star that previously had no solid
crust. Looking at the eigenfunctions, both modes bear resemblance to each
other both in the core and the outer crust, whereas they differ within the
inner crust. This is the only new mode appearing in the spectrum
(which is accessible to us, i.e. down to $\approx 18\unit{Hz}$).

Another interesting feature is the disappearance of interface modes in the
ageing star. As can be seen from Figure~\ref{fig:s-evolution}, the number of
modes in the low frequency spectrum of the star with elastic crust is smaller
than if the star was a perfect fluid. The eigenfunctions of the present modes
reveal that the elasticity of the crust prevents the large radial displacement
at the phase transitions which is characteristic for interface modes: of the
few interface modes we can find, all have their maximum radial displacement in
the thin fluid ocean, whereas their amplitude is considerably smaller within
the elastic crust.

In order to investigate this behaviour of the interface modes further, we
artificially extend the region in which the crust is elastic towards the
surface of the star while keeping the temperature profile of the star fixed at
the age of $10^6$ years. At first, we extend the crust so that only one phase
transition lies within the fluid ocean; in the low frequency spectrum, which
we investigate down to frequencies of about $13\unit{Hz}$, we are able to
identify the first 13 $g$-modes and one $i$-mode at about $16.9\unit{Hz}$; the
radial displacement of the $i$-mode is largely confined to the fluid ocean.
Next, we gradually increase the thickness of the elastic crust;
however, the spectrum does not change significantly as long as the phase
transition stays within the fluid ocean. As soon as the elastic crust extends
beyond this phase transition, the corresponding $i$-mode starts to decrease in
frequency. A further extension of the elastic crust leads to a rapid decrease
in frequency (in contrast to the stationary frequency while the phase
transition belonged to the fluid ocean, an increase in thickness of the crust
by only a few meters results in a decrease of the $i$-modes's frequency of up
to 10\,\%) and the interface mode soon vanishes from the accessible part of
the spectrum.  We conclude that the crust elasticity effectively suppresses
interface modes.  However, the numerical instabilities in the low frequency
region prevent us from quantitatively investigating at what frequencies
interface modes, which are caused by phase transitions within the elastic
crust, can be found.

\begin{figure}[ht]
    \centering
    \includegraphics[clip=true,width=0.6\textwidth]{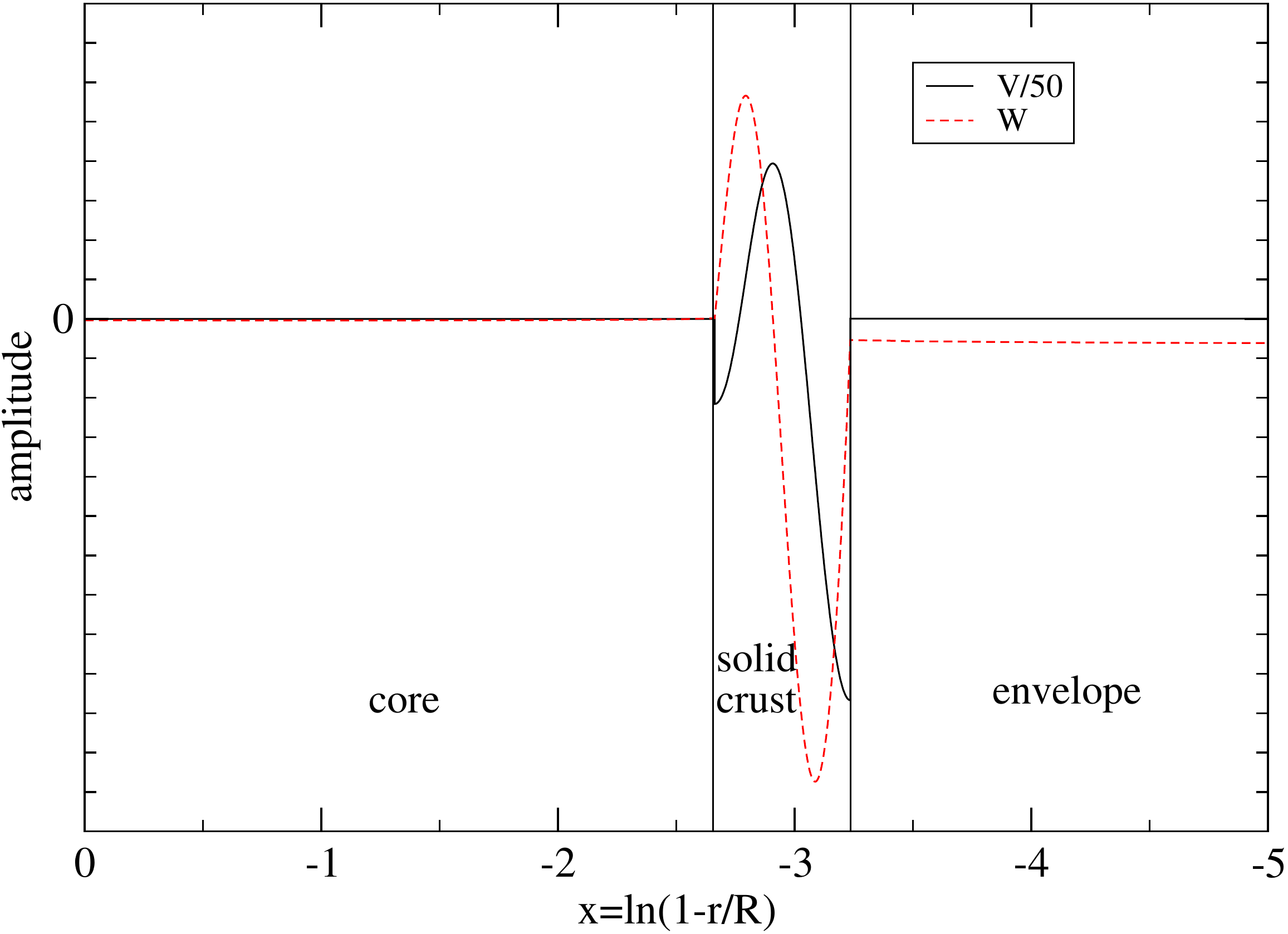}
    \caption{The displacement associated with the second shear mode, $s_2$.
    The star is four months old and its $s_2$-mode has a frequency of
    $2.588\,\unit{kHz}$.}
    \label{fig:s-mode_ef}
\end{figure}

When we consider higher frequencies, we find several narrow spikes in the
spectrum. These belong to the shear modes associated with the solid crust. We
list the frequencies of the twelve lowest shear modes for a 100 years old star
in Table \ref{tab:s-modes}.  In Figure~\ref{fig:s-mode_ef}, we show the radial
and transverse displacement of the second shear mode $s_2$ (for our neutron
star at the age of four months); we denote shear modes by $s_n$ where $n$
corresponds to the order of the oscillation mode.  Note that we have scaled
the transverse displacement, represented by the variable $V$, by a factor of
0.02.  That is, the displacement is predominantly transverse. Furthermore, the
displacement is strongly confined to the crust (apart from a small, almost
constant, radial displacement in the fluid ocean).  Both displacement
variables fit---to very good precision---one wavelength in radial
direction into the elastic crust; however, they are out of phase by
approximately $\pi / 4$ which indicates an ellipsoidal motion (rather than
circular due to the dominance of the transverse displacement) of the ``fluid
elements'' in the crust.  The eigenfunctions for the other shear modes show a
similar behaviour; higher order shear modes fit more wavelenghts into the
elastic crust, in particular, the number of wavelenghts increases by one half
per order.

\begin{table}[ht]
    \centering
    \caption{Frequencies of the high frequency modes for the neutron star
    model at the age of hundred years (as shown in Figure~\ref{fig:s-modes}).}
    \label{tab:s-modes}
    \begin{tabular}{c|c}
        Frequency & Mode \\
        {[}kHz]     & \\
        \hline
        1.304 & $s_1$ \\
        1.938 & $f$   \\
        2.205 & $s_2$ \\
        3.049 & $s_3$ \\
        3.818 & $s_4$ \\
        4.622 & $s_5$ \\
        5.504 & $s_6$ \\
        6.315 & $p_1$ \\
        6.417 & $s_7$ \\
        7.333 & $s_8$ \\
        8.229 & $s_9$ \\
        9.101 & $s_{10}$ \\
        9.489 & $p_2$    \\
        9.966 & $s_{11}$ \\
       10.844 & $s_{12}$ 
    \end{tabular}
\end{table}

The frequency of the shear waves deserves a further comment. For reasons that
will become clearer later, we again consider our neutron star at the age of
four months. It has a solid crust in the region from $R_{cc} = 10.94\unit{km}$
({\bf c}ore-{\bf c}rust interface) to $R_{co} = 11.30\unit{km}$ ({\bf
c}rust-{\bf o}cean interface). As is apparent from Figure~\ref{fig:s-mode_ef},
the $s_2$ mode essentially fits one oscillation into this region, suggesting a
wavelength of $\lambda \approx 360\unit{m}$.  This is the same behaviour as
observed for axial shear modes \cite{samuelsson_andersson_2007}. Together with
the frequency of $2.588\unit{kHz}$ we infer a wave speed of $c = \lambda f
\approx 0.93 \times 10^8 \unit{cm}/\mathrm{s}$. This is in good agreement with
the shear velocity $c_t = \sqrt{\hat{\mu}/\rho}$, which varies between $0.98 -
1.4 \times 10^8 \unit{cm}/\mathrm{s}$ in the elastic region (as calculated
from values provided by the DH EoS). The slight disagreement stems from the
fact that the $s_2$-mode does not exactly exhibit a full oscillation period.
A similar behaviour can of course also be observed in older stars, too.
However, as the elastic crust spans a wider region in older stars, the shear
speed begins to significantly vary with depth which distorts the
neat sinusoidal shape of the eigenfunctions.

In the upper panel of Figure~\ref{fig:s-evolution} we show how the high frequency part of the spectrum changes with time. We can identify the two frequencies,
$1.938\unit{kHz}$ and $6.315\unit{kHz}$, which are eigenfrequencies of the
star during its entire lifetime (apart from very short periods when a shear
mode crosses over and the two modes in question exhibit an avoided crosssing);
these are the $f$-mode and the first $p$-mode, respectively. Since the crust
does not form until the star is about 1.1 days old, we do not expect to see
other modes in the spectrum. As discussed above, we expect the frequency of
the shear modes to be determined roughly by the shear speed and the thickness
of the elastic crust; when the crust starts to crystallise, the shear modes
will have very high frequencies to begin with. Since the crystallised region
grows as the star cools, while the shear speed is subject to only small
variations, we expect the shear mode frequencies to decrease with time. The
described behaviour is clearly visible in the upper panel of
Figure~\ref{fig:s-evolution}: The shear modes appear at the top end of the
spectrum after about 1.1 days; their frequencies rapidly decrease until the
crust reaches a ``plateau'' after about 1 year. For the next hundred years or
so, the crust does not significantly increase in width (cf.
Figure~\ref{fig:crust_formation}) and hence the shear modes remain largely
unaffected.  After this phase the crust gains in width quickly, causing the
shear mode frequencies to drop almost discontinuously. After roughly 1000
years, the solid crust has almost reached its final thickness. Thus, the shear
modes do not experience a significant change in frequency later.

Our findings for stratified stars with an elastic crust are in agreement with
results from McDermott et~al.~\cite{mcdermott_etal_1988} who considered a
star with an elastic crust in Newtonian theory. Qualitatively, our numerical
calculations confirm their spectra; we find the same sets of mode, that is
$f$-, $p$-, $g$-, $i$- and $s$-modes. However, while McDermott et~al. find two
different sets of $g$-modes which they label surface $g$-modes and core
$g$-modes, according to the region to which they are mainly confined, our
simulations do not reproduce these findings as we do not have a detailed ocean
model. Furthermore, the frequencies for
the $g$-modes which McDermott et~al. find are much lower (below $3\unit{Hz}$)
than in our simulations. This is due to the rather low temperature 
($10^7\unit{K}$) used in their calculations. At these temperatures we would
expect similar frequencies in our simulations but the numerical procedure does
not allow us to investigate the spectrum at such low values.

\section{Summary and Conclusions}
\label{sec:summary}

The aim of this paper was to establish a formalism that
allows us to study neutron star oscillations in full general
relativity, accounting for as much realistic microphysics as possible.
So far we extended the standard perfect fluid formalism to account for
thermal pressure in the star's core and  derived a new set of
equations which govern perturbations in the solid crust. We provided
sample results for a particular realistic equation of state. However, our
numerical code is generic in the sense that we can easily use a different
equation of state straightaway.

We focus our attention on the low-frequency part of the oscillation spectrum
(above 15~Hz or so due to computational limitations). In this regime we find a
set of interface modes, firstly due to the (artificial) density discontinuity
at the crust-core interface and secondly there are interface modes associated
with the crust region, due to sharp changes in the low-density equation of
state.  The composition gradient shifts the frequency of the
artificial interface mode to slightly higher frequencies, and we find a set of
composition $g$-modes in the low frequency regime. The frequencies of these
$g$-modes are slightly lower than literature values from pervious
studies---this is because the composition gradient is rather small throughout
nearly the entire core; only close to the crust-core transition is the
composition gradient reasonably pronounced.

When we account for thermal pressure due to neutrons and protons in the core,
we find that a number of thermal $g$-modes enter the low frequency part
of the spectrum. Meanwhile, the high frequency interface modes are shifted
to even higher frequencies whereas the composition $g$-modes stay unaffected.
Tracking the thermal evolution of the neutron star for the first
few hundred years, we investigate how the
frequencies of the various modes evolve as the star matures through
adolescence. After approximately 10 years, all thermal $g$-modes have dropped
below $15\unit{Hz}$ and the thermal effect on the interface modes has almost
vanished entirely. During the next 100 years, as the star continues to cool,
the frequencies of the interface modes change slightly, finally leaving us
with the spectrum of a cold star.

Finally, we considered the crystallization of the crust and
investigated the associated changes in the spectrum. From our thermal
evolution, we determined  the solid region in the outer parts of the star.  We
quantified how the crust elasticity affects the interface modes associated
with the outer layers of the star and we discovered that they are suppressed
by shear stress. We discussed how the spectrum is enriched
by shear modes. As expected, our results for the shear modes are in good
agreement with results from previous work \cite{mcdermott_etal_1988}.

In summary, we have taken the first step towards a comprehensive computational
technology to study quasinormal modes of realistic compact stars. We plan
to extend the model to account for a multifluid core where the
neutrons are superfluid and the protons are superconducting. In doing so
we will consider recent equation of state data, in particular concerning the
entrainment effect, both in the core and the crust. We will also update the
cooling sequence to account for superfluid effects. We expect to report on
results in these directions in the not too distant future.

\section{Acknowledgements}

We are grateful to Ian Hawke and Ian Jones for helpful discussions. CJK
acknowledges the use of the IRIDIS High Performance Computing Facility, and
associated support services at the University of Southampton, in the
completion of this work. WCGH appreciates the use of the computer facilities
at the Kavli Institute for Particle Astrophysics and Cosmology. CJK
acknowledges financial support from the Engineering and Physical Sciences
Research Council (EPSRC) and the School of Mathematics at the University of
Southampton. NA and WCGH acknowledge support from the Science and Technology
Facilities Council (STFC) in the United Kingdom.   

\appendix

\section{Polar Perturbation Equations for the Elastic Crust}
\label{sec:app_eq_crust_polar}

The perturbations are governed by the linearised Einstein equations and the conservation laws for the matter:
\begin{equation}
    \delta G_{ab} = 8\pi \delta T_{ab}\quad\text{and}\quad
        \delta\left( \nabla_a T^{ab} \right) = 0,
\end{equation}
where $\delta T^{ab}$ denotes the \emph{total} stress-energy tensor as given
in Equation \eqref{eq:def_dT_tot}. In addition, we have definitions of the
two traction variables, $T_1$ and $T_2$, which in their expanded form read
\begin{align}
    T_1 & =  \frac{4}{3} \check{\mu}
            \left[ e^{-\lambda/2} (2l - 4) W
                   + 2r e^{-\lambda/2} \frac{\partial W}{\partial r}
                   + r^2 (K-H_2)
                   - l(l+1)V
            \right], \\
    T_2 & = -2 \check{\mu} 
            \left[ e^{\lambda/2} W
                   - (l-2)V
                   - r \frac{\partial V}{\partial r}
            \right].
\end{align}
As in the perfect fluid case
\cite{lindblom_detweiler_1983}, we  use the Lagragian variation of the
pressure as an independent variable
\begin{equation}
    X = - \frac{1}{r^l} e^{\nu/2} \Delta p
        = \frac{\gamma p}{r^2} e^{\nu/2}
            \left[
                e^{-\lambda/2} (l+1) W
                + r e^{-\lambda/2} \frac{\partial W}{\partial r}
                - r^2 (K + \frac{1}{2} H_2)
                + l(l+1)V
            \right].
\end{equation}
For brevity and clarity, we have suppressed the angular dependence in all these relations.

If we define
\begin{equation}
    Q = \frac{1}{2} r^2 e^{-\lambda} \nu' \ ,
    \end{equation}
    and
    \begin{equation}
            m = \frac{1}{2} (l+2) (l-1) \ ,
\end{equation}
and make liberal use of Equations \eqref{eq:tov_lambda} - \eqref{eq:def_M},
the full set of perturbation equations for the elastic crust reads:
\begin{align}
    \frac{\partial H_1}{\partial r}
        & = \left[ \frac{1}{2}(\lambda' - \nu')
                     - \frac{l+1}{r} \right]
              H_1
            + \frac{e^{\lambda}}{r}
                \left[ H_2 + K - 16 \pi (\rho + p) V \right]
                                                        \label{eq:odeH1} \\
    \frac{\partial K}{\partial r}
        & = \frac{1}{r} H_2
            + \frac{m+1}{r}H_1
            + \left( \frac{1}{2} \nu' - \frac{l+1}{r} \right) K
            - \frac{8 \pi (\rho + p) e^{\lambda/2} }{r} W,
                                                        \label{eq:odeK} \\
    \frac{\partial H_0}{\partial r}
        & = + \frac{\partial K}{\partial r}
            -r e^{-\nu} \omega^2 H_1
            - \left( \frac{1}{2} \nu'
                + \frac{l-1}{r} \right) H_0
            - \left( \frac{1}{2} \nu'
                + \frac{1}{r}   \right) H_2
            + \frac{l}{r} K
            - \frac{16 \pi}{r} T_2,
                                                        \label{eq:odeH0} \\
    \frac{\partial W}{\partial r}
        & = - \frac{l+1}{r} W
            + r e^{\lambda/2} \left[
                \frac{e^{-\nu/2}}{\gamma p} X
                - \frac{l(l+1)}{r^2} V
                + \frac{1}{2} H_2
                + K
            \right],
                                                        \label{eq:odeW} \\
    \frac{\partial V}{\partial r}
        & = \frac{1}{2 \check{\mu} r} T_2
            + \frac{e^{\lambda/2}}{r} W
            + \frac{2-l}{r} V,
                                                        \label{eq:odeV} \\
    \frac{\partial T_2}{\partial r}
        & = - \frac{1}{2} r e^{\lambda} (\rho + p) H_0
            + r e^{\lambda - \nu / 2}
              \left( X 
                - \frac{1}{2r^2} e^{\nu / 2} T_1 
              \right)
            + \left[
                \frac{1}{2}(\lambda' - \nu')-\frac{l+1}{r}
              \right] T_2
                                                        \nonumber \\
        & \quad + \left[
                    \frac{4 m e^{\lambda} \check{\mu}}{r}
                    -e^{\lambda-\nu} r \omega^2 (\rho + p) 
                  \right] V
                + e^{\lambda / 2} p' W.
                                                        \label{eq:odeT2}
\end{align}
In addition to these six ordinary differential equations (ODEs), we have three
algebraic relations:
\begin{align}
    H_2 
        & = H_0 + 64 \pi \check{\mu} V,
                                                        \label{eq:alg1} \\
    \left( 2M + Q + mr \right) H_0
        & = 8\pi r^3 e^{-\nu/2} X
            - \left[ (m+1)Q - \omega^2 r^3 e^{-(\lambda+\nu)} \right] H_1
                                                        \nonumber \\
        & \quad + \left[ mr - \omega^2 r^3 e^{-\nu} 
                     - \frac{e^{\lambda}}{r} Q (2M + Q - r) \right] K
                                                        \label{eq:alg2} \\
        & \quad + 8 \pi r T_1 - 16 \pi r e^{-\lambda} T_2,
                                                        \nonumber \\
    \frac{2}{3} e^{-\nu/2} \check{\mu} r^2 X - \frac{1}{4} \gamma p T_1
        & = \check{\mu} \gamma p
            \left[ 2 e^{-\lambda/2} W - r^2 K + l(l+1) V \right].
                                                        \label{eq:alg3}
\end{align}

The origin of the equations are (for brevity, we denote the
components of the Einstein equations with $[ab]$ as a shortcut for
$G_{ab} = 8\pi T_{ab}$): Equations
\eqref{eq:odeH1}, \eqref{eq:odeK} and \eqref{eq:odeH0} are $[t\theta]$,
$[tr]$ and $[r\theta]$, respectively;
\eqref{eq:odeW} and \eqref{eq:odeV} are due to the definitions of $X$ and
$T_2$, respectively; \eqref{eq:odeT2} is $\delta \left( \nabla_a
T^a_{\theta} \right) = 0$;
\eqref{eq:alg1} is the difference $[\theta\theta]-[\phi\phi]$;
\eqref{eq:alg2} is $[rr]$ and \eqref{eq:alg3} is obtained by
removing $\partial_r W$ from the definitions of $X$ and $T_1$.

\section{Low-Frequency Equations}
\label{sec:app_eq_low_freq}

In the low-frequency domain, we need to circumvent numerical difficulties
arising from an algebraic equation in the `standard formulation' of the
perfect fluid problem. This led us to using $V$ instead of $X$ as an
independent variable. The perturbation equations for this set of variables
then take the form
\begin{align}
    \frac{\partial H_1}{\partial r}
        & = \left[
                + \frac{1}{2} \left( \lambda' - \nu' \right)
                - \frac{l+1}{r}
            \right] H_1 
            + \frac{e^{\lambda}}{r}
              \left[ H_0 + K -16 \pi \left( \rho + p \right) V \right],
                                                    \label{eq:pfV_odeH1} \\
    \frac{\partial K}{\partial r} 
        & = \frac{1}{r} H_0
            + \frac{m+1}{r} H_1
            + \left( \frac{1}{2} \nu' - \frac{l+1}{r} \right) K
            - \frac {8 \pi \, \left( \rho  + p \right)}{r} e^{\lambda/2} W,
                                                    \label{eq:pfV_odeK} \\
    \frac{\partial W}{\partial r}
        & = - \left( \frac{l+1}{r} + \frac{p'}{F} \right) W
            + r e^{\lambda/2}
              \left[
                \frac{ \rho + p }{F}
                    \left( e^{-\nu} \omega^2 V + \frac{1}{2} H_0 \right)
                - \frac{l(l+1)}{r^2} V
                + \frac{1}{2} H_0
                + K
              \right],
                                                    \label{eq:pfV_odeW} \\
    \frac{\partial V}{\partial r}
        & = \left( \frac{p'}{F} - \frac{\rho'}{\rho+p}
                + \nu' - \frac{l}{r} \right) V
            + \frac{1}{2 \omega^2} e^{\nu}
                \left(
                    \frac{p'}{F} - \frac{\rho'}{\rho + p}
                \right)
                \left( H_0 + \frac{1}{r} e^{-\lambda/2} \nu' W \right)
                                                                \nonumber \\
        & \quad + r H_1
            - \frac{1}{r} e^{\lambda/2} W,
                                                    \label{eq:pfV_odeV}
\end{align}
where we have introduced the abbreviation
\begin{equation}
    F = \gamma p
        + n_\mathrm{n} \frac{\partial p_{th}^{\text{n}}}{\partial n_\mathrm{n}}
        + n_\mathrm{p} \frac{\partial p_{th}^{\text{p}}}{\partial n_\mathrm{p}}.
\end{equation}
The Einstein equations imply the following algebraic relation between the
perturbation variables, which we use in order to calculate $H_0$:
\begin{align}
    \left[ 3M - 4 \pi r^3 \rho + m r
    \right] H_0
        & = - \left[
                (m+1) Q - \omega^2 r^3 e^{-(\lambda+\nu)}
            \right] H_1
                                                        \nonumber \\
        & \quad + \left[
                mr - \omega^2 r^3 e^{-\nu}
                - \frac{e^{\lambda}}{r} Q (2M + Q - r)
            \right] K
                                                        \label{eq:pfV_alg1} \\
        & \quad + 8 \pi e^{\lambda/2} Q (\rho + p) W
            + 8 \pi r^3 e^{-\nu} \omega^2 \left( \rho + p \right) V,
                                                        \nonumber
\end{align}
The origin of the equations is basically the same as in previous work, but still deserves a few comments. In order to arrive at our set of
equations, we start from the Detweiler \& Lindblom set of equations \cite{detweiler_lindblom_1985}.
Since the algebraic
relation~\eqref{eq:pfluid_V} causes numerical problems, we solve it for $X$
and substitute  into the other equations. After some algebraic manipulations
and making use of the background equations as well as the other perturbation
equations, we then arrive at our set of perturbation equations. 

Equations \eqref{eq:pfV_odeH1}, \eqref{eq:pfV_odeK} and \eqref{eq:pfV_alg1} follow from
$[t\theta]$, $[tr]$ and $[rr]$ of the perturbed Einstein equations, respectively; \eqref{eq:pfV_odeW} is due to
the definition of (the no longer appearing) $X$; \eqref{eq:pfV_odeV} is
$\delta \left( \nabla_a T^a_r \right) = 0$ and \eqref{eq:pfluid_V},
which has been used as an auxiliary equation only, is $\delta \left(
\nabla_a T^a_{\theta} \right) = 0$.

Before we can solve Equation \eqref{eq:pfV_alg1} for $H_0$, we have to assure
that its coefficient is non-singular at any point. This is easy to see, as we
have
\begin{equation}
    3M = 3 \int_0^r 4 \pi r'^2 \rho(r') dr'
        \ge 3 \int_0^r 4 \pi r'^2 \rho(r) dr'
        = 4 \pi r^3 \rho,
\end{equation}
where we used Equation \eqref{eq:tov_lambda} for the first equality and
monotonicity of the density, $\rho(r') \ge \rho(r)$ for $r' \le r$, inside the
star for the estimate. Thus, the coefficient will be singular only at the
origin, $r=0$, where we use a Taylor expansion anyway.

\subsection{Behaviour at the Origin}

Since the perturbation equations (see Equations \eqref{eq:pfV_odeH1} -
\eqref{eq:pfV_alg1})
are singular at the origin due to the use of spherical coordinates, we 
expand all variables by Taylor series near the origin, $r=0$. For all our
perturbation variables $H_0$, $H_1$, $K$, $V$ and $W$, we use an
expansion of the form
\begin{equation}
    \label{eq:taylorQ}
    Q(r) = Q_0 + \frac{1}{2} r^2 Q_2 + \mathcal{O} \left( r^4 \right).
\end{equation}
The field equations imply that the first order corrections vanish. The zeroth
order constraints imposed by the perturbation equations are
\begin{subequations}
\label{eq:tayl0}
\begin{align}
    H_{10}
        & = \frac{ 2l K_0 + 16 \pi \left( \rho_0 + p_0 \right) W_0}{ l(l+1) }, 
        &
    V_0
        & = - \frac{1}{l} W_0,
        &
    H_{00}
        & = K_0,
\end{align}
\end{subequations}
which demonstrates that, once $K_0$ and $W_0$ are chosen, the remaining
expansion coefficients, $H_{00}$, $H_{10}$ and $V_0$ are determined.
Here and in the following, we need the expansion coefficients of the
background quantities, $\nu$, $\lambda$, $p$ and $\rho$ as well. We 
expanded them in the same way as specified in Equation \eqref{eq:taylorQ};
their second order coefficients are given by
\begin{align}
    \lambda_2 & = \frac{16}{3} \pi \rho_0, \\
    \nu_2     & = \frac{8}{3} \pi \left( \rho_0 + 3 p_0 \right), \\
    p_2       & = - \frac{1}{2} \nu_2 \left( \rho_0 + p_0 \right), \\
    \rho_2    & = \frac{ p_2 \left( \rho_0 + p_0 \right) }{ \gamma p_0 }.
\end{align}
The second order coefficients of the perturbation variables are then given by
the following linear system:

\begin{subequations}
\begin{align}
    \lefteqn{- \frac{l+3}{2} H_{12}
    + \frac{1}{2} K_2
    - 8 \pi (\rho_0 + p_0) V_2
    + \frac{1}{2} H_{02} }
                                                        \nonumber \\
        & \qquad \qquad \qquad \qquad
          = \frac{1}{2} (\nu_2 - \lambda_2) H_{10}
            - \lambda_2 K_0
            + 8 \pi
                \left[
                    \rho_2 + p_2 + \lambda_2 (\rho_0 + p_0)
                \right] V_0,
                                                        \\
    \lefteqn{\frac{m+1}{2} H_{12}
    - \frac{l+3}{2} K_2
    - 4 \pi (\rho_0 + p_0) W_2
    + \frac{1}{2} H_{02}}
                                                        \nonumber \\
        & \qquad \qquad \qquad \qquad
          = - \frac{1}{2} \nu_2 K_0
            + 2 \pi
                \left[
                    2 (\rho_2 + p_2) + \lambda_2 (\rho_0 + p_0)
                \right] W_0,
                                                        \\
    \lefteqn{- (m+1) F_0 V_2
        - \frac{l+3}{2} F_0 W_2}
                                                        \nonumber \\
        & \qquad \qquad \qquad \qquad
          = - \frac{1}{2} \left( 3 F_0 + p_0 + \rho_0 \right) K_0
            + \left[ \frac{m+1}{2} \lambda_2 F_0 - l p_2 
            - e^{-\nu_0} \omega^2 (\rho_0 + p_0) \right] V_0,
                                                        \\
    \lefteqn{ (l+2) V_2 + W_2 }
        & \qquad \qquad \qquad \qquad
          = 2 H_{10}
            - \frac{e^{\nu_0}}{\omega^2}
              \left(
                \frac{\rho_2}{\rho_0+p_0} - \frac{p_2}{F_0}
              \right) K_0
                                                        \nonumber \\
        & \qquad \qquad \qquad \qquad
            + \left[
                \frac{l}{2} \lambda_2 + 2 \nu_2
                + \left(
                    2 - \frac{ l e^{\nu_0} \nu_2 }{ \omega^2 }
                  \right)
                  \left(
                    \frac{ p_2 }{ F_0 } - \frac{\rho_2}{\rho_0 + p_0}
                  \right)
              \right] V_0,
                                                        \\
    \lefteqn{\frac{m}{2} K_2
        - \frac{m}{2} H_{02}}
        & \qquad \qquad \qquad \qquad
          = \left(
                \frac{m+1}{2} \nu_2 - \omega^2 e^{-\nu_0}
            \right) H_{10}
            + \left(
                \omega^2 e^{-\nu_0} - \frac{1}{2} \nu_2
            \right) K_0
                                                        \nonumber \\
        & \qquad \qquad \qquad \qquad
            - 8 \pi e^{-\nu_0} \omega^2 (\rho_0 + p_0) V_0
            + 8 \pi p_2 W_0.
                                                        \label{eq:pfY_t6}
\end{align}
\end{subequations}

\section{Numerical Strategy - General Case}
\label{sec:app_numerics}

Here, we describe the details of the numerical procedure for
the interior solution. In this work, we consider a star with three layers.
However, as more physics are taken into account, we will have to slice the neutron
star into more layers, each of which extends over a region in which the
neutron star matter is homogeneous in the sense that it can be described by
the same set of equations. Thus, in anticipation of extending our work to
account for superfluidity, where we will certainly have more than three
distinct regions within the star, we consider a star with $n$ layers. The same procedure with
exactly the same underlying idea and very similar junction conditions has
already been explained in great detail for a three-layer star by Lin
et~al.~\cite{lin_etal_2008}. Hence, using this $n$-layer description and
applying it to our 3-layer star together with the junction conditions from
Section~\ref{sec:perturbations} is straightforward.

As an example, consider a neutron star which consists of $n$ layers. The
interfaces between the layers, including the center and the surface of the
star, are at the radii $0 = R_0 < R_1 < R_2 < \cdots <
R_{n-1} < R_n = R$, where $R$ is the star's radius. Within each of the different
layers, the dynamics and the perturbations, respectively, are governed by a
certain set of equations which obviously depends on the nature of the matter
within this region. It could be, for instance, a perfect fluid, the elastic
crust or a superfluid. We write the equations for layer $i$
as
\begin{equation}
    \label{eq:layer_i_ode}
    \frac{d{}^{(i)}\mathbf{Y}}{dr}
        = {}^{(i)}\mathbf{Q} \cdot {}^{(i)}\mathbf{Y}
        \quad\text{for}\quad r \in [R_{i-1},R_i],
\end{equation}
where ${}^{(i)}\mathbf{Y} = \left( y_1, \ldots, y_{k_i} \right)$ is an abstract
vector field with as many entries, say $k_i$, as there are independent
variables in this layer (in most cases, we have four or six
independent variables); ${}^{(i)}\mathbf{Q} = {}^{(i)}\mathbf{Q} (r,l,
\omega)$ is a $k_i \times k_i$ matrix (depending also on the background
fields, which we have suppressed in this notation). The variables $y_i$ are
placeholders for the corresponding perturbations variables, like $H_1$, $K$,
$V$, etc.

Let us now find the general solution in layer $i$. Since we, a priori, do not
have any boundary conditions for this layer, we choose some set of
values for ${}^{(i)} \mathbf{Y} (R_{i-1})$ and integrate through this layer
using Equation \eqref{eq:layer_i_ode} until we reach $r = R_i$. In order to
find the general solution, we repeat this procedure $k_i$ times; each
time starting with a different set of values for ${}^{(i)} \mathbf{Y}
(R_{i-1})$, where all these `start vectors' ought to be linearly independent.
This generates $k_i$ linearly independent solutions, ${}^{(i)} \mathbf{Y}_j
(r)$ (with $j = 1, \ldots, k_i$), and the general solution
is a linear combination of these, i.e.
\begin{equation}
    {}^{(i)} \mathbf{Y} (r)
        = \sum_{j=1}^{k_i} c_{i,j} {}^{(i)} \mathbf{Y}_j (r)
        \quad\text{for}\quad r \in [R_{i-1},R_i],
\end{equation}
where the coefficients $c_{i,j}$ with $i = 1, \ldots, n$ (denoting the layer)
and $j = 1, \ldots, k_i$ (counting the different solutions inside a particular
layer) are constants to be determined by interface and boundary conditions. 
Before we turn to these, we mention some peculiarities while
integrating the different layers. Firstly, in some cases it is possible to
reduce the computational effort. This happens if some interface condition
essentially is a fixed boundary condition for this layer; for instance, a
condition could require a variable to vanish at an interface and we will then,
of course, apply this condition when choosing the `start vectors', effectively
reducing the number of linearly independent solutions. Secondly, the innermost
layer (number 1 in our way of counting) needs special treatment as the
differential equations, due to the use of spherical coordinates, are most
likely singular at the origin, $r = 0$. We will therefore expand the solutions
at the origin into Taylor series, calculate the Taylor coefficients up to
second order and use this expansion up to a small radius at which the ODEs can
safely be integrated numerically. Thirdly, the outermost layer $n$ is
traditionally integrated from the surface of the star inwards since in most of
the cases the Lagrangian pressure perturbation, $X$, is used as a variable
which has to vanish at the surface, $X(R) = 0$.

Finally, we use the (remaining) interface conditions in order to determine the
$\sum_i k_i$ coefficients $c_{i,j}$. An interface condition connects the
solutions between layer $i-1$ and $i$. After fixing the overall normalisation
by choosing the value of one of the coefficients $c_{i,j}$, we expect to have
$\left( \sum_i k_i \right) - 1$ interface and boundary conditions (less the
conditions we used to reduce the computational effort)
in order to uniquely determine the solution. The actual interface conditions
can take a variety of forms and a general discussion would go well beyond the
scope and necessity; therefore, we restrict ourselves to give two common
examples only. First, some variables like the metric perturbations are usually
continuous across interfaces, i.e. $[H_1]_{R_i} = 0$, whereas other
variables simply vanish, i.e. $T_2(R_i^-) = 0$. The superscript `$-$' in $R_i^-$
is to denote that we consider the limit of $T_2$ at the inner edge of the
interface at radius $R_i$. These exemplary interface
conditions read, respectively
\begin{align}
    \left[ H_1 \right]_{R_i}
        & = 0 \qquad\Leftrightarrow
        &
    \sum_{j=1}^{k_i} c_{i,j} {}^{(i)} H_1(R_i)
        & = \sum_{j=1}^{k_{i+1}} c_{i+1,j} {}^{(i+1)} H_1(R_i),            \\
    T_2(R_i^-)
        & = 0 \qquad\Leftrightarrow
        &
    \sum_{j=1}^{k_i} c_{i,j} {}^{(i)} T_2(R_i)
        & = 0.
\end{align}

\subsection{The  Boundary Conditions}

The problem we are solving involves matching the solution across  different
interfaces. The simplest case involves the switch from the equations for low
frequencies to the formulation of Detweiler \& Lindblom
\cite{detweiler_lindblom_1985}.  Obviously, all the variables ($H_0$, $H_1$,
$K$, $V$, $W$ and $X$) have to be continuous at the matching point. Since only
four of them are independent, we may freely choose four out of these six. It
turns out (by inspection) that it is numerically advantageous to use the set
of $H_0$, $K$, $V$ and $X$ for matching inside the core ($[H_0]_{R_V} =
[K]_{R_V} = [V]_{R_V} = [X]_{R_V} = 0$ in short notation). By choosing those
four variables, we find that the condition number of the resulting matrix is some orders of
magnitude lower than for other combinations. There are also a couple of other
sets for which the matrix has a lower condition number but we are unable to
reason this behaviour nor are we able to find a pattern. As a rule of thumb,
the variables $H_0$ and $K$ should be amongst the four chosen ones.

The actual physical interfaces are those which confine the elastic crust. As
explained in Section~\ref{sec:perturbations}, we have five
continuous variables at both sides of the elastic crust, namely $H_0$, $H_1$,
$K$, $W$ and $T_2$. We have to use all of them; however, at the core-crust
interface, we use the continuity of $T_2$, and put it in the form $T_2 = 0$ in
order to reduce the number of independent solutions in the crust by one, down
to five. Formally, we use $[H_0]_{R_{cc}} = [H_1]_{R_{cc}} = [K]_{R_{cc}} =
[W]_{R_{cc}} = [H_0]_{R_{co}} = [H_1]_{R_{co}} = [K]_{R_{co}} = [W]_{R_{co}} =
0$ and $T_2(R_{co}^-) = 0$, where $R_{cc}$ and $R_{co}$ are the radii of the
core-crust and the crust-ocean interfaces, respectively.

At the center of the star and at the surface, we impose the boundary
conditions from Detweiler \& Lindblom \cite{detweiler_lindblom_1985}.

\bibliographystyle{apsrev}
\bibliography{references}

\end{document}